\documentclass[pra,aps,amsmath,amssymb,amsfonts,onecolumn,nofootinbib]{revtex4}
\usepackage{amssymb}
\usepackage{}
\usepackage{bm,mathrsfs}

\usepackage{graphicx}
\usepackage{epsfig}
\usepackage{amsmath,bbm}
\usepackage{amsfonts,amssymb}
\usepackage{times}
\usepackage{verbatim}
\usepackage[sort&compress]{natbib}

\usepackage{amsmath}
\usepackage[colorlinks,breaklinks,linkcolor=blue,anchorcolor=blue,citecolor=blue,urlcolor=blue,
dvipdfm
]{hyperref}

\newcommand{\be}{\begin{equation}}
\newcommand{\ee}{\end{equation}}
\newcommand{\bea}{\begin{eqnarray}}
\newcommand{\eea}{\end{eqnarray}}

\def\>{\rangle}
\def\<{\langle}

\def\qed{\leavevmode\unskip\penalty9999 \hbox{}\nobreak\hfill
     \quad\hbox{\leavevmode  \hbox to.77778em{%
               \hfil\vrule   \vbox to.675em%
               {\hrule width.6em\vfil\hrule}\vrule\hfil}}
     \par\vskip3pt}
     
\begin{document}

\newtheorem{theorem}{Theorem}
\newtheorem{lemma}[theorem]{Lemma}
\newtheorem{corollary}[theorem]{Corollary}
\newtheorem{proposition}[theorem]{Proposition}
\newtheorem{definition}[theorem]{Definition}
\newtheorem{example}[theorem]{Example}
\newtheorem{conjecture}[theorem]{Conjecture}
\title{Noise-induced distributed entanglement in atom-cavity-fiber system}

\author{Dong-Xiao Li}
\affiliation{Center for Quantum Sciences and School of Physics, Northeast Normal University, Changchun, 130024, People's Republic of China}
\affiliation{Center for Advanced Optoelectronic Functional Materials Research, and Key Laboratory for UV Light-Emitting Materials and Technology
of Ministry of Education, Northeast Normal University, Changchun 130024, China}

\author{Xiao-Qiang Shao\footnote{Corresponding author: shaoxq644@nenu.edu.cn}}
\affiliation{Center for Quantum Sciences and School of Physics, Northeast Normal University, Changchun, 130024, People's Republic of China}
\affiliation{Center for Advanced Optoelectronic Functional Materials Research, and Key Laboratory for UV Light-Emitting Materials and Technology
of Ministry of Education, Northeast Normal University, Changchun 130024, China}

\author{Jin-Hui Wu}
\affiliation{Center for Quantum Sciences and School of Physics, Northeast Normal University, Changchun, 130024, People's Republic of China}
\affiliation{Center for Advanced Optoelectronic Functional Materials Research, and Key Laboratory for UV Light-Emitting Materials and Technology
of Ministry of Education, Northeast Normal University, Changchun 130024, China}

\author{X. X. Yi}
\affiliation{Center for Quantum Sciences and School of Physics, Northeast Normal University, Changchun, 130024, People's Republic of China}
\affiliation{Center for Advanced Optoelectronic Functional Materials Research, and Key Laboratory for UV Light-Emitting Materials and Technology
of Ministry of Education, Northeast Normal University, Changchun 130024, China}
\date{\today}

\begin{abstract}
{The distributed quantum computation plays an important role in large-scale quantum information processing. In the atom-cavity-fiber system, we put forward two efficient proposals to prepare the steady entanglement of two distant atoms with dissipation. The atomic spontaneous emission and the loss of fiber are exploited actively as powerful resources, while the effect of cavity decay is inhibited by quantum Zeno dynamics and quantum-jump-based feedback control. These proposals do not require precisely tailored Rabi frequencies or coupling strength between cavity and fiber. Furthermore, we discuss the feasibility of extending the present schemes into the systems consisting of  two atoms at the opposite ends of the $n$ cavities connected by $(n-1)$ fibers, and the corresponding numerical simulation reveals that a high fidelity remains achievable with current experimental parameters.}
\end{abstract}

\maketitle

\section{Introduction}
Distributed quantum computation (DQC) is a large cluster consisting of many small or medium scale quantum computers to run large scale quantum operation \cite{1704.02620ref27}. The concept of DQC was first suggested by Grover \cite{Compref12}, and Cleve \textit{et al.} \cite{Compref3}, and  the theoretical research of DQC can be broadly split into two basic types: (i) Solve the problems where classical distributed computing are incapable. For instance, Tani \textit{et al.} \cite{Compref20} and D'Hondt \textit{et al.} \cite{Compref5} proposed quantum algorithms to exactly solve the leader election problem in
anonymous networks. (ii)  Simulate large-capacity quantum computer by several small-capacity quantum computers. For example, Cirac \textit{et al.} \cite{pra4249} performed quantum computations nonlocally among distant nodes in a quantum network.

Quantum entanglement \cite{pra032323ref6,pra022317ref1}, as one of the most intriguing features in the quantum world, reflects that the states of composite systems cannot be translated into product states of subsystems. It lies at the heart of quantum information processing (QIP) and has been widely applied in various fields, such as superdense coding \cite{pra032323ref9,pra060305ref3},
teleportation \cite{pra032323ref10,pra060305ref5,pra060305ref6}, quantum fingerprinting \cite{pra060305ref7,pra060305ref8}, and direct characterization of quantum dynamics \cite{pra060305ref9}.
The generation of quantum entanglement by distributed quantum computation is crucial in large-scale quantum information processing \cite{prl050501ref1,prl050501}. Up to present, several schemes have been suggested for generating quantum entanglement distribution
over long distance \cite{prl5242,pra052305re3,pra052305re1,pra052305re4,pra052305ref5,pra052305,prl070501ref2,prl010503,prl070501ref3,prl070501ref4,prl070501,1704.02620ref81,1704.02620ref166,1704.02620ref46}.
Particularly, Pellizzari proposed a scheme to transfer quantum information between two atoms
via an optical fiber in the presence of decoherence, where they performed an adiabatic passage by keeping two cavities in their respective vacuum states {during the whole period} \cite{prl5242}. Serafini \textit{et al.} investigated the realization of effective quantum gates between two atoms in distant cavities coupled by an optical fiber, where the highly reliable swap and entangling gates were achieved \cite{prl010503}. In addition, a variety of experiments demonstrate the feasibility of distant entanglement with high fidelities \cite{prl070501ref6,prl070501ref7,prl070501ref8}. However, the success rate for preparation of distant entanglement will be decreased because of the quantum dissipation, and generally a precisely tailored Rabi frequencies or coupling strength between cavity and fiber is required.

It is acknowledged that the quantum dissipation originates from the {weak} interaction between quantum systems and its environment. The dissipation results in the decoherent effect, which works against the interests of QIP tasks, and is inevitable at the actual situation. Fortunately, the utility value of dissipation has been discovered recently, which represents a dramatic breakthrough on quantum computation. Different schemes were put forward to prepare the entangled states by dissipation \cite{pra022329ref14,njp.11.083008,pra054302ref1,myZnref15,myZnref16,Shenepl,myZnref20,myZnref18,shankar2013,PhysRevA.90.054302,PhysRevA.92.033403,Jin:17,myZnref21,myZnref22}, \textit{e.g.} Kastoryano \textit{et al.} suggested a dissipative scheme to create entangled states of two atoms in an optical cavity, which is more effective than the schemes based on unitary dynamics \cite{pra054302ref1}. Shen \textit{et al.} generalized the ideas of Kastoryano's and generate distant steady-state entanglement of two atoms trapped in separate cavities directly coupled to each other \cite{myZnref15} or connected by a dissipative bosonic medium \cite{Shenepl}.
{  Jin \textit{et al.} utilized the combined effect of the unitary dynamics and dissipative process to prepare a distributed steady-state entanglement \cite{Jin:17}. Nevertheless, there remains some potential problems in the previous schemes, for example, either the qubit need to be driven by fields with well-chosen frequencies or the spontaneous emission may decrease the fidelity of the target state. Furthermore, the schemes of distributed entanglement only take advantage of one resonant non-local mode and sacrifice other non-resonant modes in a multi-cavity system.}

The quantum Zeno effect \cite{myZnref23} can be used to compete deoherence remarkably, which hinders the initial state of a quantum system evolving into other states by repetitively frequent observation. After the effect was certified in several experiments \cite{myZnref24,myZnref25}, a more general process, the quantum Zeno dynamics was raised, where the system can evolve away from its initial state under a multidimensional projection \cite{myZnref28,myZnref30,myZnref31,myZnref32}. The quantum Zeno dynamics provides us with the possibility to significantly restrain the cavity decay in QIP tasks \cite{PhysRevA.80.062323,shaojpb,myZnref35,PhysRevA.89.033856,PhysRevA.91.012325,myZnref36,myZnref37,PhysRevLett.117.140502,Li:17}.
{ On the other hand, the quantum feedback is usually applied in decoherence suppression, entanglement production, and entanglement protection \cite{pra022332ref14,pra022332ref12,pra022332ref15,pra022332ref13,PhysRevA.76.010301,PhysRevA.78.012334,PhysRevA.84.022332,pra022332ref18,pra032307}. This technology is based on immediately feeding back the measurement results to the quantum system of interest and then alter its subsequent dynamics. For instance, Mancini \textit{et al.} controlled steady state Einstein-Podolsky-Rosen correlations for two bosonic modes interacting via parametric Hamiltonian by the quantum feedback \cite{pra022332ref13}. Carvalho \textit{et al.} proposed a direct feedback based on quantum-jump detection to achieve the maximal entangled states between two atoms in a cavity \cite{PhysRevA.76.010301}, and then they found that the quantum-jump-based feedback can protect highly entangled states against decoherence by comparing the effects of different control Hamiltonians and detection processes \cite{PhysRevA.78.012334}.}

{ In this paper, we propose two dissipative schemes to generate distant entangled Bell state and Knill-Laflamme-Milburn (KLM) state of two atoms with high fidelity, respectively, where two $\Lambda$ atoms are trapped in two separated cavities connected by a fiber. We eliminate the unconcerned states by dispersive microwave fields, and inhibit the cavity decay by the quantum Zeno dynamics and the quantum-jump-based feedback
control. The schemes have four distinctive features: (i) The target state is independent of the concrete initial states. (ii) The schemes are robust against the loss of photon, and the spontaneous emission of atoms and the decay of fibers are powerful resources to create target states. (iii) The atoms in different cavities are more convenient to be manipulated, and needn't precisely tailored Rabi frequencies or coupling strength between cavity and fiber. (iv) The schemes can be directly generalized into a system, that $n$ cavities are connected by $(n-1)$ fibers, without introducing the non-local modes.}

\section{The full and effective Markovian master equations of system to generate entangled states}\label{II}
\subsection{The generation of the Bell state}
\begin{figure}
\centering
\includegraphics[scale=0.14]{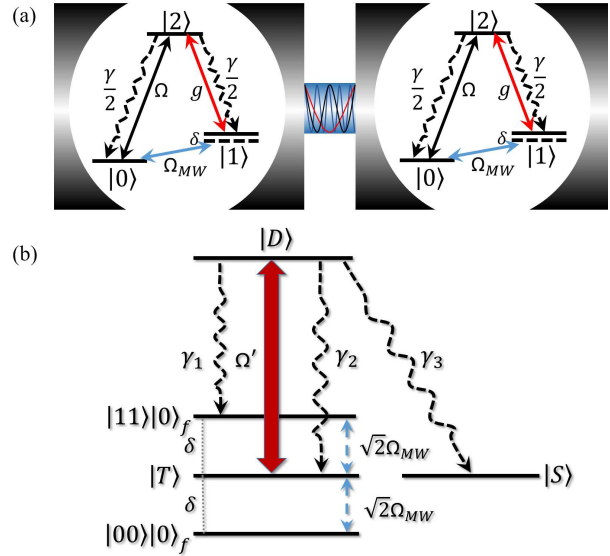}
\caption{\label{model}(a) Diagrammatic illustration of the cavity-atom-fiber system and the atomic level for preparation of the Bell state $|S\rangle=(|01\rangle-|10\rangle)|0\rangle_f/\sqrt{2}$. (b) The effective transitions of the reduced system.}
\end{figure}

Figure \ref{model}(a) illustrates two $\Lambda$ atoms trapped in two separated optical cavities connected by a fiber which supports $N$ modes. {The target state is $|S\rangle=(|01\rangle-|10\rangle)|0\rangle_f/\sqrt{2}$, where $|0\rangle_f$ denotes the vacuum states of fiber.} Each atom has an excited state $|2\rangle$, and two ground states $|0\rangle$ and $|1\rangle$. A quantized cavity field resonantly drives the transition between states $|1\rangle$ and $|2\rangle$ with coupling strength $g$. A laser resonantly drives the transition between states $|0\rangle$ and $|2\rangle$ with Rabi frequency $\Omega$. The two ground states are coupled by a microwave field with Rabi frequency $\Omega_{MW}$, detuned by $\delta$. In the interaction picture, the Hamiltonian reads
\begin{eqnarray}\label{B3CH}
H_I&=&H_C^I+H_Q^I,\\
H^I_C&=&\sum_{i=1}^2\Omega_{MW}|1\rangle_i\langle0|+{\rm H.c.}+\delta|1\rangle_i\langle1|-\delta a_i^\dag a_i+\sum_m^N(\Delta_m-\delta) b_m^\dag b_m,\nonumber\\
H_Q^I&=&\sum_{i=1}^{2}a_i|2\rangle_i\langle1|g+\Omega_i|2\rangle_i\langle0|+\sum_m^NJa_{i}^\dag b_m+{\rm H.c.},\nonumber
\end{eqnarray}
where $a_i~(a_i^\dag)$ and $b_m~(b_m^\dag)$ denote the annihilation (creation) operators of the $i$th quantized cavity and the $m$th fiber mode, respectively. $J$ is the coupling strength between the fiber and the cavity. $\Delta_m$ is the frequency
difference between the $m$th fiber mode and the cavity
mode. {  In the short fiber limit, the number of modes in fiber, which are resonant with the cavity modes, satisfies $N=l\kappa_f/(2\pi c)\leq1$, \textit{i.e.}, we can only consider one mode in fiber is resonant with the cavity mode and the other modes are ignored because of the large detuning \cite{prl5242,prl010503}. $l,~\kappa_f$ and $c$ denote the
length of fiber, the decay rate of fiber and the speed of
light, respectively.} Thus $H_C^I$ and $H_Q^I$ can be simplified as
\begin{eqnarray}\label{HCI}
H^I_C&=&\sum_{i=1}^2\Omega_{MW}|1\rangle_i\langle0|+{\rm H.c.}+\delta(|1\rangle_i\langle1|-a_i^\dag a_i-b^\dag b),\\
H_Q^I&=&\sum_{i=1}^{2}a_i|2\rangle_i\langle1|g+\Omega_i|2\rangle_i\langle0|+Ja_{i}^\dag b+{\rm H.c.}.
\end{eqnarray}

The full master equation of present system can be expressed as
\begin{eqnarray}\label{3cbmaster}
\dot\rho=-i[H_I,\rho]+\sum_j L_j\rho L_j^\dag-\frac{1}{2}(L_j^\dag L_j\rho+\rho L_j^\dag L_j),
\end{eqnarray}
where the Lindblad operators $L_j$ are
\begin{eqnarray}\label{3CLind}
L_\gamma^{1(2)}=\sqrt{\frac{\gamma}{2}}|0(1)\rangle_1\langle2|,\  L_\gamma^{3(4)}=\sqrt{\frac{\gamma}{2}}|0(1)\rangle_2\langle2|,\
L_\kappa^{i}=\sqrt{\kappa_i}a_i,\
L_{\kappa_f}=\sqrt{\kappa_f}b,
\end{eqnarray}
with $\gamma$ and $\kappa_{i(f)}$ being the rates of atomic decay and cavity (fiber) decay, respectively.

For simplicity, we assume $\Omega_1=\Omega_{2}=\Omega$ in what follows. Referring to the quantum Zeno dynamics \cite{myZnref31,myZnref32}, we can write the $H_Q^I$ as
\begin{eqnarray}\label{limit}
H_Q^I=\Omega(H_c+KH),
\end{eqnarray}
where $H_c$ is interaction between atoms and classical field. Without loss of generality, we first set $J\geq g$, then we have $K=g/\Omega$ and $H=H_g+(J/g)H_J$, where $H_{g(J)}$ represents the interaction between the cavity filed and atoms (fiber). While for $J<g$, we can also obtain a similar Hamiltonian where the roles of $g$ and $J$ interchange. The detailed derivation has been shown in Appendix \ref{A}. In the limiting condition, $K\rightarrow\infty$, \textit{i.e.} $g\gg\Omega$, the term $H^I_Q$ can be rewritten as
\begin{eqnarray}\label{HQIDT}
H_Q^I&=&\frac{\sqrt{2}\Omega}{\sqrt{G_2+2}}|D\rangle\langle T|+{\rm H.c.},\\
|D\rangle&=&\frac{1}{\sqrt{G_2+2}}\Big(|21\rangle|0\rangle_f+|12\rangle|0\rangle_f-\frac{g}{J}|11\rangle|1\rangle_f\Big),\\
|T\rangle&=&\frac{1}{\sqrt{2}}(|01\rangle+|10\rangle)|0\rangle_f,
\end{eqnarray}
where $G_2=(g/J)^2$, the subscript ``$2$'' represents the number of cavities, and $|1\rangle_f$ indicates the single photon state of fiber. We have omitted the vacuum state of cavity mode, since it is decoupled to our system in the regime of quantum Zeno dynamics. Then the effective Hamiltonian can be rewritten as
\begin{eqnarray}\label{B3CHeff}
H_{\rm{eff}}&=&\Omega'|D\rangle\langle T|+\sqrt{2}\Omega_{MW}|T\rangle(\langle00|+\langle11|){}_f\langle0|+{\rm H.c.}\nonumber\\
&&+\delta(2|11\rangle\langle11|\otimes|0\rangle_f\langle0|+|T\rangle\langle T|+|S\rangle\langle S|+|D\rangle\langle D|),
\end{eqnarray}
where $\Omega'=\sqrt{2}\Omega/\sqrt{G_2+2}$.

In order to analytically derive an effective master equation, here we suppose $\gamma=\kappa_f$ for simplicity, then the effective Lindblad operators (See Appendix \ref{B}) can be denoted by
\begin{eqnarray}\label{B3CLind}
L_{\rm{eff}}^{1}=\gamma_1|11\rangle|0\rangle_f\langle D|,\  \  L_{\rm{eff}}^{2(3)}=\gamma_{2(3)}|S(T)\rangle\langle D|,
\end{eqnarray}
where $\gamma_1=\sqrt{\gamma(G_2+1)/(G_2+2)}$, and $\gamma_{2(3)}=\sqrt{\gamma/(2G_2+4)}$.
The corresponding effective Markovian master equation is
\begin{eqnarray}\label{3cbeffmaster}
\dot\rho=-i[H_{\rm eff},\rho]+\sum_{k=1}^3 L_{\rm eff}^k\rho L_{\rm eff}^{k\dag}-\frac{1}{2}(L_{\rm eff}^{k\dag} L_{\rm eff}^k\rho+\rho L_{\rm eff}^{k\dag} L_{\rm eff}^k).
\end{eqnarray}

On the basis of Eq.~(\ref{3cbeffmaster}), we plot the effective transitions of quantum states in Fig.~\ref{model}(b) {to show the physics principle of the preparation of our target state}. There are four ground states $|00\rangle|0\rangle_f,|11\rangle|0\rangle_f,|T\rangle,|S\rangle$ and one excited state $|D\rangle$. The ground states $|00\rangle|0\rangle_f,|11\rangle|0\rangle_f$ and $|T\rangle$ can translate to each other by the microwave field. The excited state $|D\rangle$ are coupled with the ground state $|T\rangle$ by a laser with Rabi frequency $\Omega'$ and can decay into states $|11\rangle|0\rangle_f, |T\rangle$, and $|S\rangle$ because of the atomic spontaneous emission and the decay of the fiber. Due to the detuning of microwave field, the state $|S\rangle$ becomes the unique steady state of system. Combining the above requirements, for arbitrary initial state, the system will always be stable at the desired state $|S\rangle$.

\subsection{The generation of the KLM state}
\begin{figure}
\centering
\includegraphics[scale=0.14]{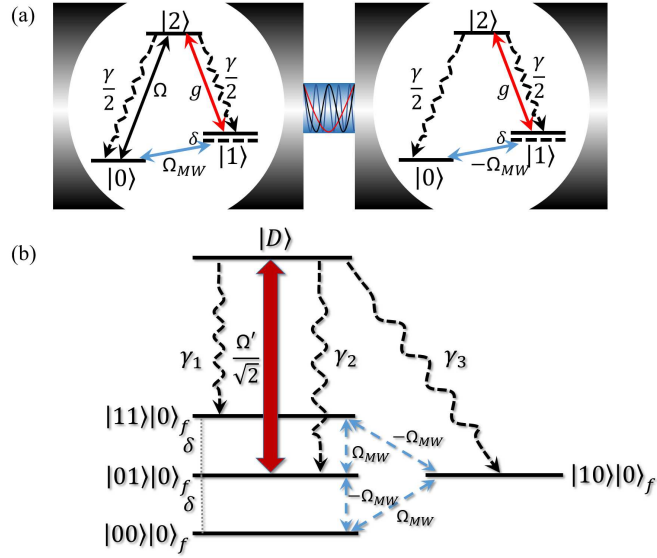}
\caption{\label{klmmodel}(a) Diagrammatic illustration of the cavity-atom-fiber system and the atomic level for preparation of the bipartite KLM state $|K_1\rangle=(|00\rangle+|10\rangle+|11\rangle)|0\rangle_f/\sqrt{2}$. (b) The effective transitions of the reduced system.}
\end{figure}
{{ The KLM state was first introduced by Knill, Laflamme, and Milburn \cite{nature409}, which plays important roles in quantum information processing, quantum state teleportation and error correction \cite{pra012321ref19,pra012321}. The success probability of the scalable quantum computation can be notably improved by the KLM state \cite{nature409}. Based on this advantage, Franson \textit{et al.} designed a high-fidelity quantum logic operations \cite{prl137901} and put forward a more general form of KLM state by elementary linear-optics gates and solid-state approach \cite{cheng2012ref13}.} In our scheme, we consider the bipartite KLM state as $|K_1\rangle=(|00\rangle+|10\rangle+|11\rangle)|0\rangle_f/\sqrt{3}$ and we have three modifications from previous setup to generate the bipartite KLM state: (i) Remove the laser with Rabi frequency $\Omega$ of the second atom. (ii) Add to a $\pi$ phase difference of the microwave field acting on two atoms. (iii) Set $\delta=|\Omega_{MW}|$. The diagram is plotted in Fig.~\ref{klmmodel}(a). The effective transitions are shown in Fig.~\ref{klmmodel}(b). Similar to the derivation as aforementioned by quantum Zeno dynamics, the full and effective Hamiltonian can be obtained as
\begin{eqnarray}\label{klm3CHeff}
H_I&=&H_C^I+H_Q^I,\\
H^I_C&=&\Omega_{MW}(|1\rangle_1\langle0|-|1\rangle_2\langle0|)+{\rm H.c.}+\sum_{i=1,2}\delta|1\rangle_i\langle1|-\delta a_i^\dag a_i-\delta b^\dag b,\nonumber\\
H_Q^I&=&\sum_{i=1}^{2}a_i|2\rangle_i\langle1|g+Jba_{i}^\dag+\Omega_1|2\rangle_1\langle0|+{\rm H.c.},\nonumber
\end{eqnarray}
and
\begin{eqnarray}\label{klm3CHeff1}
H_{\rm{eff}}&=&\frac{\Omega'}{\sqrt{2}}|01\rangle|0\rangle_f\langle D|+\Omega_{MW}(|00\rangle-|11\rangle)(\langle10|-\langle01|)\otimes|0\rangle_f\langle0|+{\rm H.c.}\nonumber\\
&&+\delta(2|11\rangle\langle11|+|01\rangle\langle 01|+|10\rangle\langle 10|)\otimes|0\rangle_f\langle0|+\delta|D\rangle\langle D|,
\end{eqnarray}
where $\Omega'=\sqrt{2}\Omega/\sqrt{G_2+2}$. The corresponding effective Lindblad operators are
\begin{eqnarray}\label{klm3CLind}
L_{\rm{eff}}^{1(2)}=\gamma_{1(2)}|10(01)\rangle|0\rangle_f\langle D|,\  \   L_{\rm{eff}}^{3}=\gamma_3|11\rangle|0\rangle_f\langle D|,
\end{eqnarray}
where $\gamma_{1,2,3}$ are the same as those in Eq.~(\ref{B3CLind}).
From the Fig.~\ref{klmmodel}(b), we can find the four ground states $|00\rangle|0\rangle_f,|10\rangle|0\rangle_f,|01\rangle|0\rangle_f$ and $|11\rangle|0\rangle_f$ are coupled by the microwave fields. The state $|01\rangle|0\rangle_f$ can be pumped into the excited state $|D\rangle$ by a laser with Rabi frequency $\Omega'/\sqrt{2}$. Utilizing the dissipation of the atoms and fiber, the excited state further decay into states $|10\rangle|0\rangle_f, |01\rangle|0\rangle_f$, and $|11\rangle|0\rangle_f$. Meanwhile, the presence of the detuning of
microwave field, which satisfies $\delta=\Omega_{MW}$, causes the target state $|K_1\rangle$ to be the unique steady state of system. Ultimately, all of the conditions stabilize the system into the KLM state $|K_1\rangle$.}

\section{The influences of relevant parameters}\label{III}
\subsection{For the Bell state}
\begin{figure*}
\begin{minipage}[t]{0.32\linewidth}
\centering
\includegraphics[scale=0.32]{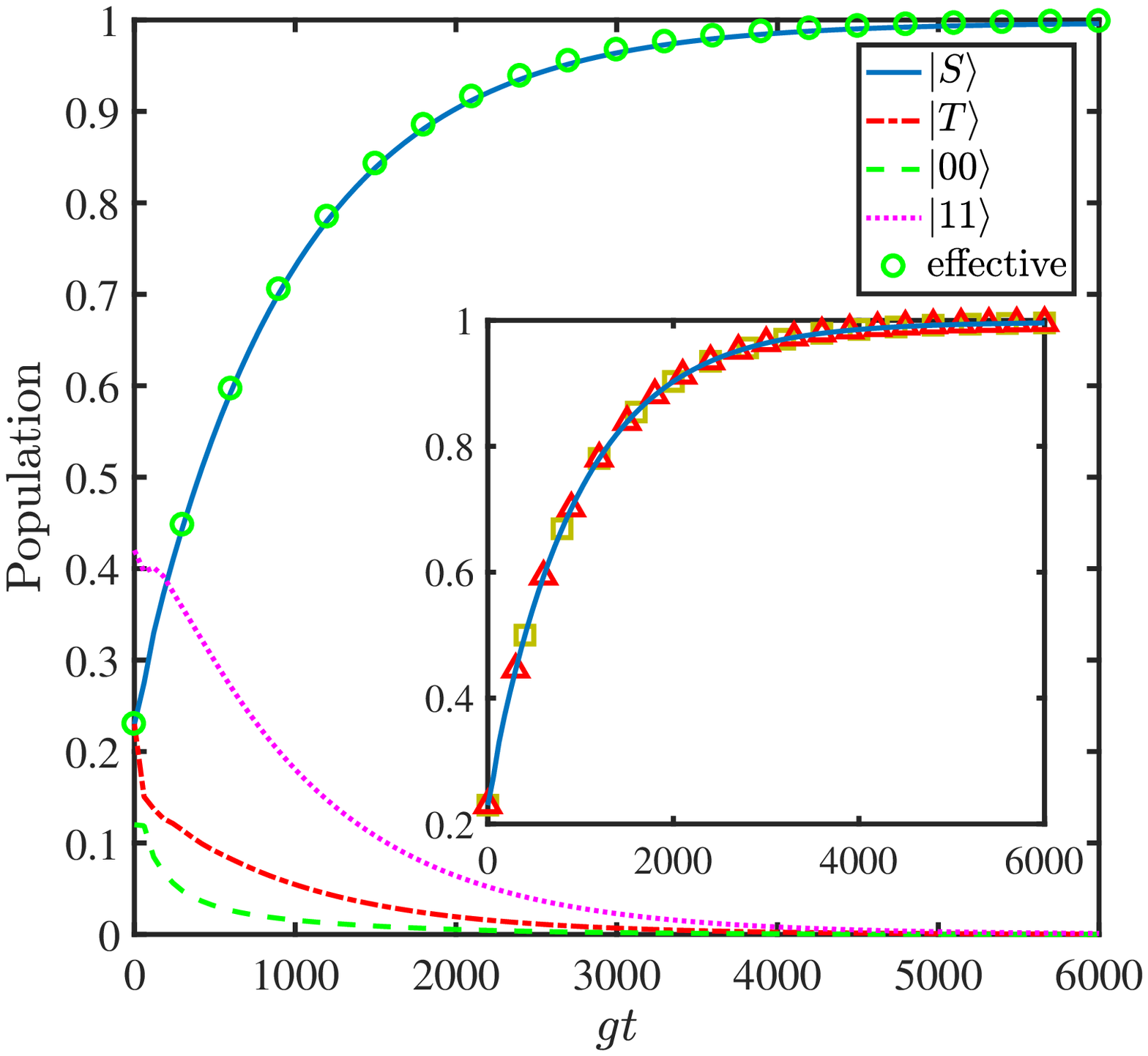}
 \centerline{(a)}
\end{minipage}
\begin{minipage}[t]{0.32\linewidth}
\centering
\includegraphics[scale=0.32]{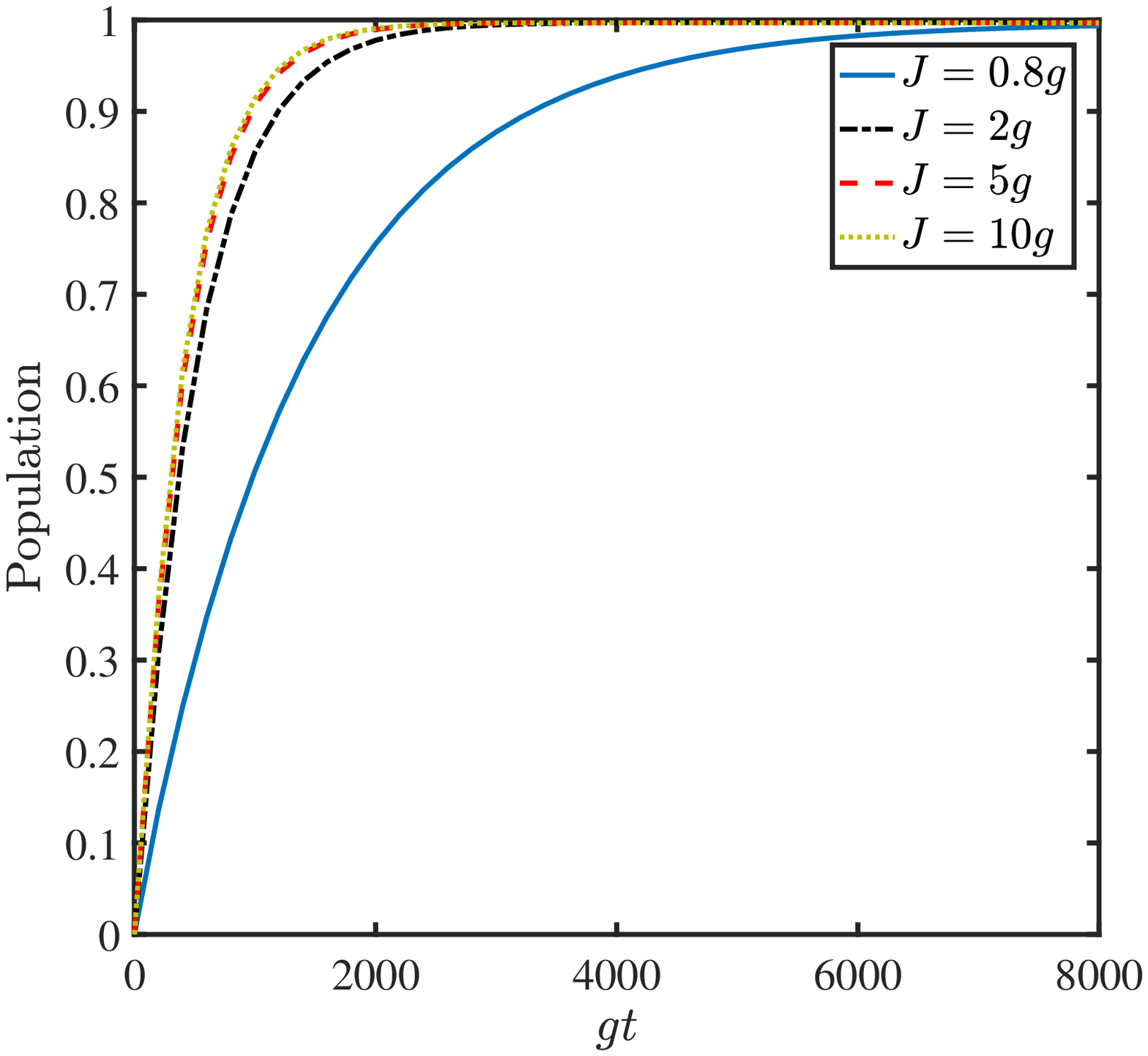}
 \centerline{(b)}
\end{minipage}
\begin{minipage}[t]{0.32\linewidth}
\centering
\includegraphics[scale=0.32]{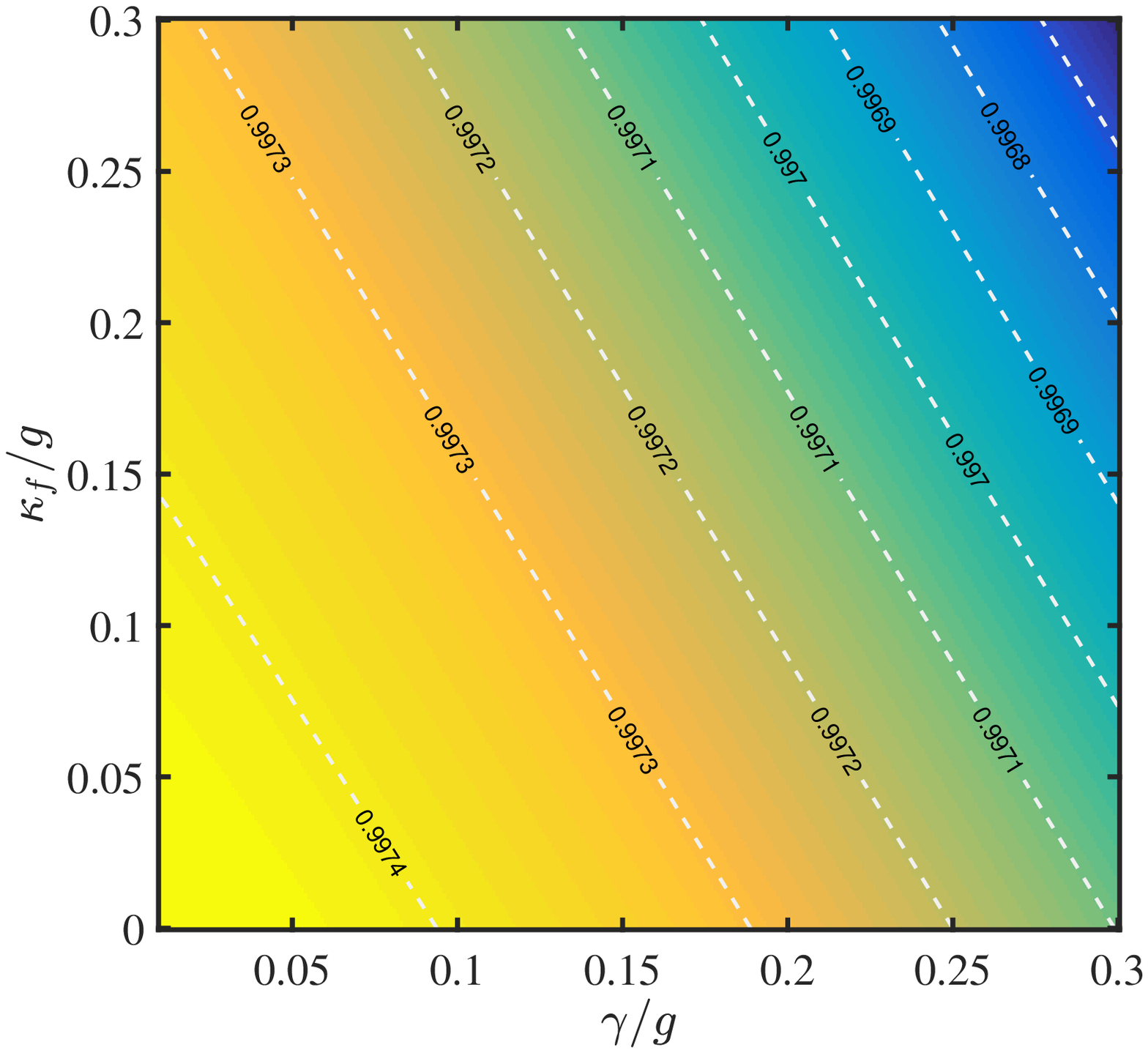}
 \centerline{(c)}
\end{minipage}
\caption{\label{fulleffect} (a) The populations as functions of $gt$ governed by full and effective master equations. The initial states are $\rho_0=(a|00\rangle\langle00|+b|11\rangle\langle11|+c|10\rangle\langle10|+d|01\rangle\langle01|)\otimes|0\rangle_f\langle0|$, where $a=0.12$, $b=0.42$, $c=0.35$, and $d=0.11$. We set $\gamma=\kappa_f=0.1g$ and $J=g$. { The inset compares the populations of $|S\rangle$ between single excitation (solid line), two excitations (empty square) and three excitations (empty triangle).} (b) The populations of state $|S\rangle$ with different coupling strengths between the fiber and
the cavity $J$. The initial states are chosen as $|00\rangle|0\rangle_f$ and set $\gamma=\kappa_f=0.1g$. (c) Contour plot (dashed lines) of the populations of $|S\rangle$ in the steady states with perfect cavities and $J=g$. The other parameter are all set as $\Omega=0.05g,\Omega_{MW}=0.3\Omega,\delta=0.05g$ and $\kappa_1=\kappa_2=0$.}
\end{figure*}

{In order to confirm the above analysis, we also plot the time evolution of populations of states in Fig.~\ref{fulleffect}(a). The inset compares the populations of $|S\rangle$ between single excitation (solid line), two excitations (empty square) and three excitations (empty triangle), where the population of state $|m\rangle$ is defined as $P=\langle m|\rho(t)|m\rangle$. Since the three lines are in good agreement with each other, for simplicity, we only consider the situation with single excitation in the other figures. In the Fig.~\ref{fulleffect}(a), the populations of $|S\rangle$ (solid line), $|T\rangle$ (dotted-dashed line), $|00\rangle$ (dashed line) and $|11\rangle$ (dotted line) are governed by the full master equation of Eq.~(\ref{3cbmaster}). We randomly initialize the state at $\rho_0=(a|00\rangle\langle00|+b|11\rangle\langle11|+c|10\rangle\langle10|+d|01\rangle\langle01|)\otimes|0\rangle_f\langle0|$, where $a=0.12$, $b=0.42$, $c=0.35$, and $d=0.11$. The population of target state $|S\rangle$ can be $99.6\%$ at $gt=6000$. and the population of $|S\rangle$ governed by the effective master equation is illustrated by the empty circle, which is in agreement with the solid line and convincingly demonstrates the validity of the reduced system. Therefore, we can forecast the behavior of the full master equation by the reduced system.}

In Fig.~\ref{fulleffect}(b), we investigate the influences of the coupling strength $J$ on the preparing process of target state. Starting from the initial state $|00\rangle|0\rangle_f$, the populations of state $|S\rangle$ are plotted as functions of $gt$ with different coupling strengths  between the fiber and the cavity $J$. { Compared with previous schemes \cite{pra054302ref1,myZnref15}, our scheme can generate the high population of target state without special requirement of Rabi frequencies or coupling strength between cavity and fiber. The coupling strength just alters the convergence time, but does not affect the final quality of entanglement.} The stronger $J$ is given, the shorter convergence time. The behavior can be understood from Fig.~\ref{model}(b). The increasing of $J$ depresses the effective decay rate $\gamma_1$, but enhances the laser with Rabi frequency $\Omega'$ and the effective decay rate $\gamma_{2(3)}$, where $\gamma_1=\sqrt{\gamma[1+(g/J)^2]/[2+(g/J)^2]}$, $\Omega'=\sqrt{2}\Omega/\sqrt{(g/J)^2+2}$, and $\gamma_2=\gamma_3=\sqrt{\gamma/[2(g/J)^2+4]}$. So the convergence time of $|S\rangle$ will be shortened while the $J$ increases.
\begin{figure*}
\begin{minipage}[t]{0.49\linewidth}
\centering
\includegraphics[scale=0.43]{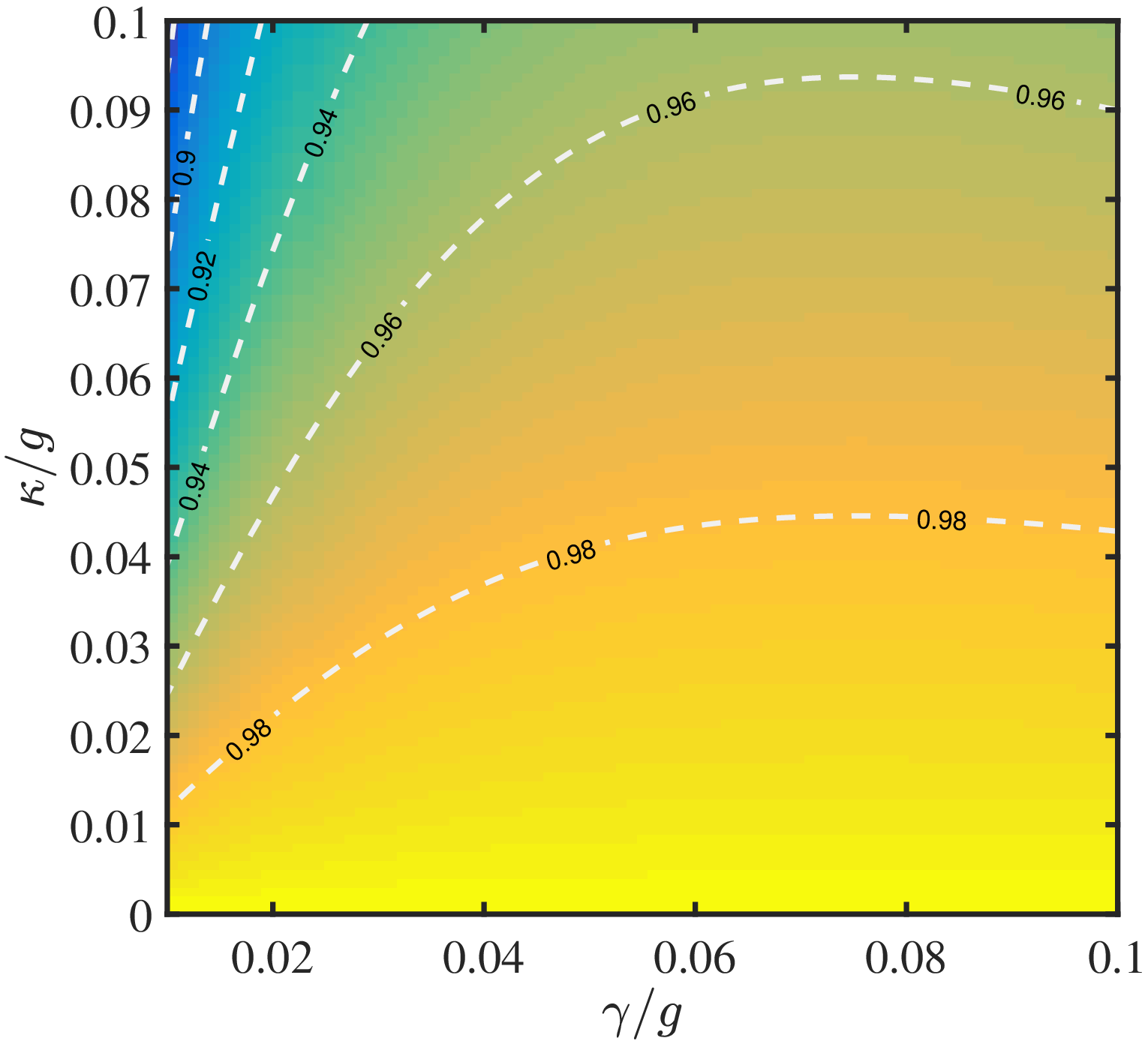}
 \centerline{(a)}
\end{minipage}
\begin{minipage}[t]{0.49\linewidth}
\centering
\includegraphics[scale=0.43]{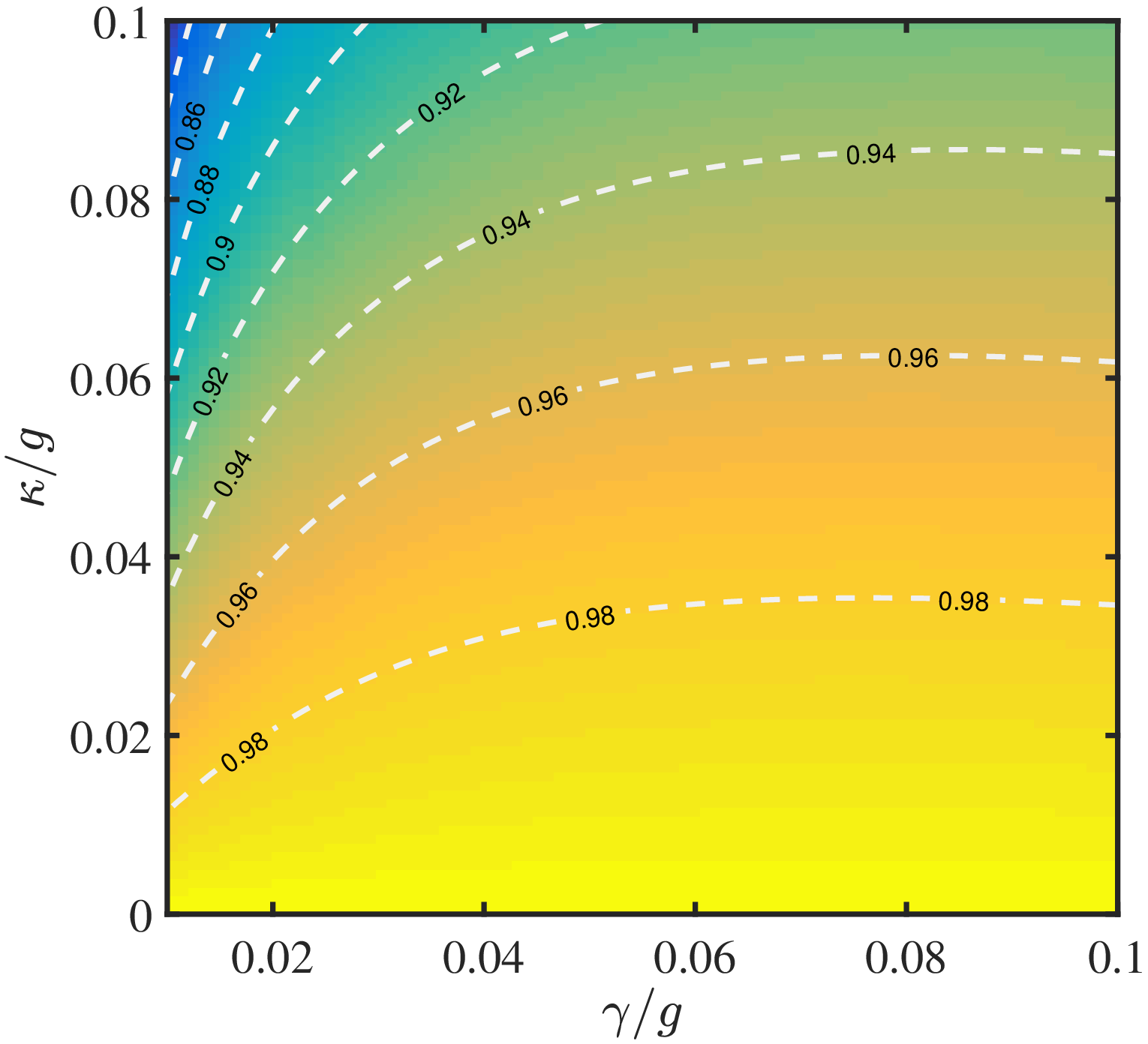}
 \centerline{(b)}
 \end{minipage}
\caption{\label{k123} Contour plot (dashed lines) of the populations of $|S\rangle$ in the steady states with leaky cavities. We set (a) $\kappa_f=0$ and (b) $\kappa_f=\kappa$. The other parameters are both $\Omega = 0.01g,\Omega_{MW} = 1.4\Omega,\delta =0.017g,J = g,$ and $\kappa_1=\kappa_2=\kappa.$ }
\end{figure*}

\begin{figure*}
\begin{minipage}[t]{0.49\linewidth}
\centering
\includegraphics[scale=0.43]{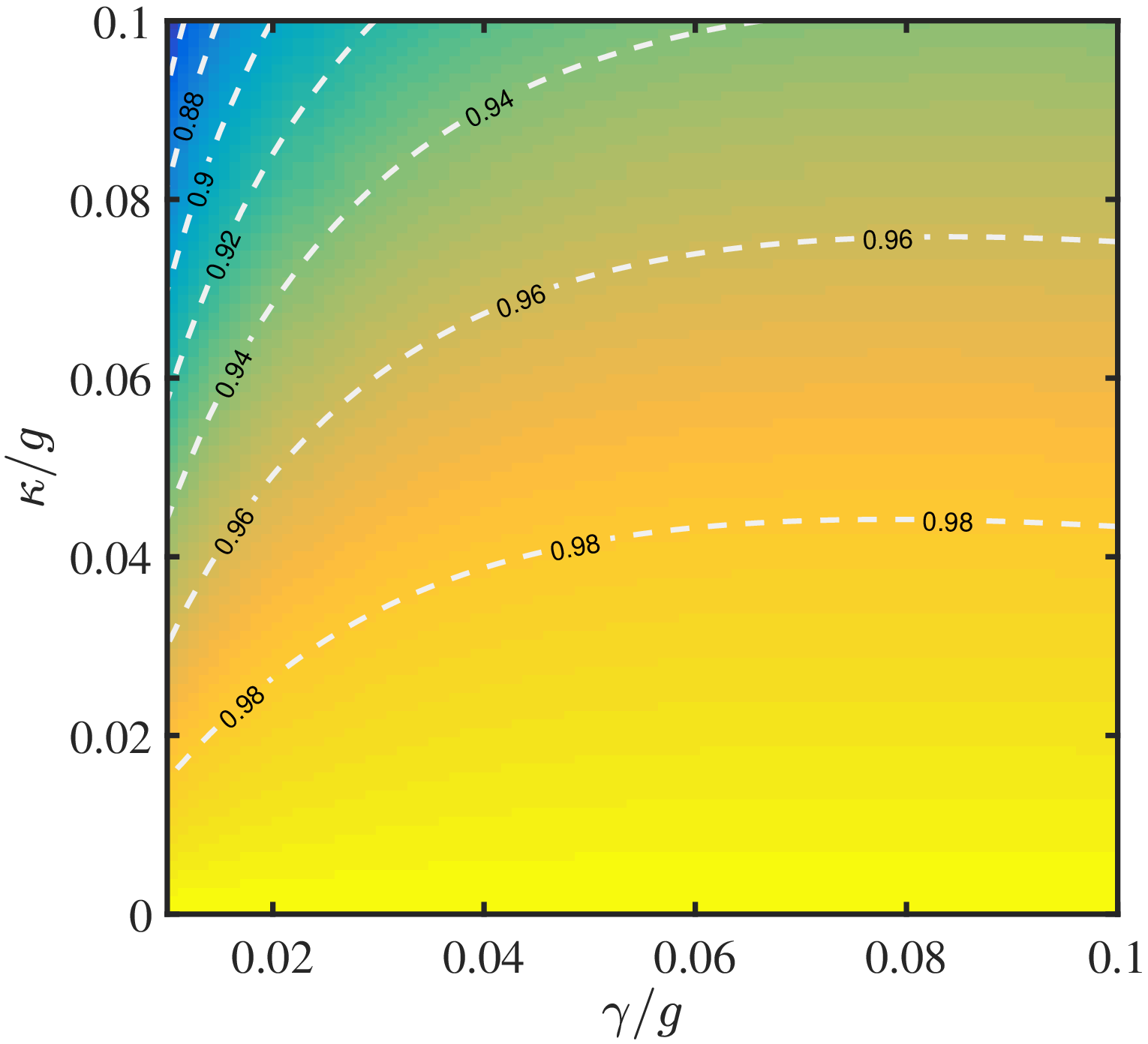}
 \centerline{(a)}
\end{minipage}
\begin{minipage}[t]{0.49\linewidth}
\centering
\includegraphics[scale=0.43]{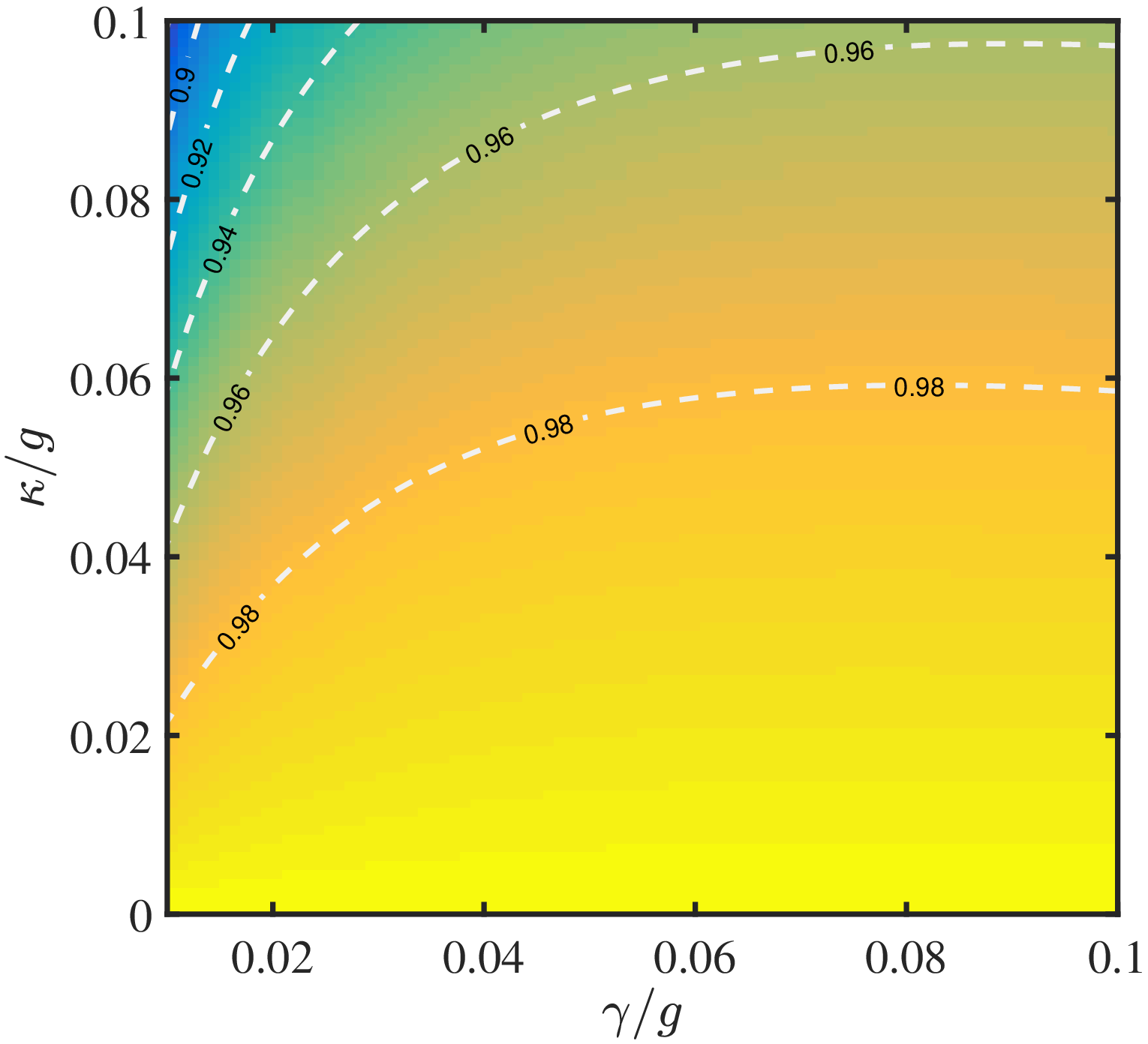}
 \centerline{(b)}
 \end{minipage}
\caption{\label{feedback} (a) Contour plot (dashed lines) of the populations of $|S\rangle$ in the steady states based on the feedback master equation with the detection of the first cavity. (b) Contour plot (dashed lines) of the populations of $|S\rangle$ in the steady states based on the feedback master equation with the detection of the two cavities. The feedback parameters are both $\eta=0.5\pi$.   The other parameters: $\Omega = 0.01g,\Omega_{MW} = 1.4\Omega,\delta =0.017g,J = g,$ and $\kappa_1=\kappa_2=\kappa_f=\kappa$. }
\end{figure*}

In Fig.~\ref{fulleffect}(c), we study the relations of atomic spontaneous emission rate, fiber decay rate, and the populations of state $|S\rangle$ solved by $\dot\rho=0$ with full master equations. The straight dashed lines are the contours of the
populations and indicate the effects of atomic and fiber decay are equivalent. We can obtain the population is still above $99.67\%$ even if the atomic and fiber decay rate both arrive at $0.3g$. Then we show the contours of populations of steady target state in leaky cavities in Fig.~\ref{k123}(a) and Fig.~\ref{k123}(b). In Fig.~\ref{k123}(a) we consider the fiber is perfect and the population of $|S\rangle$ can be beyond $90\%$ at the region $0.014g<\gamma<0.1g$ and $\kappa=0.1g$. Furthermore, for a leaky fiber in Fig.~\ref{k123}(b), even though $0.03g<\gamma<0.1g$ and $\kappa=0.1g$, the population of $|S\rangle$ is still above $90\%$. The result fully justifies that the present scheme can generate distant entangled state with a wide range of decoherence parameters.

\begin{figure}
\centering
\includegraphics[scale=0.26]{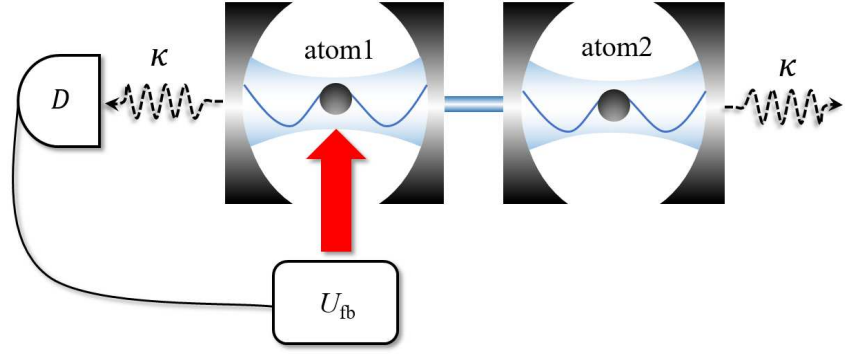}
\caption{\label{feedmodel} Diagram of the quantum-jump-based feedback control. The decay of the first cavity, $\kappa_1=\kappa$, is detected by a photodetector $D$,
triggering the feedback operation $U_{\rm fb}$ acting on the first atom.}
\end{figure}

However, comparing Fig.~\ref{k123}(a) and Fig.~\ref{k123}(b) with Fig.~\ref{fulleffect}(c), we find that the decay of cavities remain adverse. {In order to further prevent the cavity decay, we introduce the quantum-jump-based feedback control, which was proposed by Carvalho \textit{et al.} \cite{PhysRevA.76.010301} and had a variety of implementations to control decoherence in QIP tasks \cite{PhysRevA.76.010301,PhysRevA.78.012334,PhysRevA.84.022332}. By monitoring the environment and feeding back with
a suitable interaction, the master equation can be modified to revise the state of system into the target state. { Considering the current scheme, we schematically show the process of quantum-jump-based feedback control in Fig.~\ref{feedmodel}. The feedback operator $U_{{\rm fb}}=\exp{[-i\eta(|0\rangle_1\langle1|+|1\rangle_1\langle0|)\otimes{\rm I_2}]}$ acts on the first atom right after a photodetector $D$ detecting the decay of the first cavity.} The corresponding feedback master equation can be written as
\begin{eqnarray}\label{feedsingle}
\dot\rho=-i[H_I,\rho]+(\mathcal{L}_d+\kappa\mathcal{D}[U_{{\rm fb}}a_1]+\kappa\mathcal{D}[a_2])\rho,
\end{eqnarray}
where
$
\mathcal{L}_d\rho$ incorporates the dissipation of atomic decay and fiber decay, and
$\mathcal{D}[a]\rho=a\rho a^\dag-(a^\dag a\rho+\rho a^\dag a)/2.
$

In Fig.~\ref{feedback}(a), we plot the contour of the population of steady target state governed by Eq.~(\ref{feedsingle}). The region of population greater than $90\%$ has been enlarged into $0.02g<\gamma<0.1g$ and $\kappa=0.1g$ by the feedback control. Furthermore, we consider the feedback control immediately after the detection event of the two cavities and replace the last two terms of Eq.~(\ref{feedsingle}) by $\kappa\sum_{i=1}^2\mathcal{D}[U_{{\rm fb}}a_i]$. The corresponding contour of the population of $|S\rangle$ in the steady state is illuminated in Fig.~\ref{feedback}(b). When $0.0125g<\gamma<0.1g$ and $\kappa=0.1g$, the population can be above $90\%$. The result is better than that in Fig.~\ref{k123}(a), where we consider the perfect fiber. These performance notably proves the experimental feasibility of our scheme.

\begin{figure}
\centering
\includegraphics[scale=0.51]{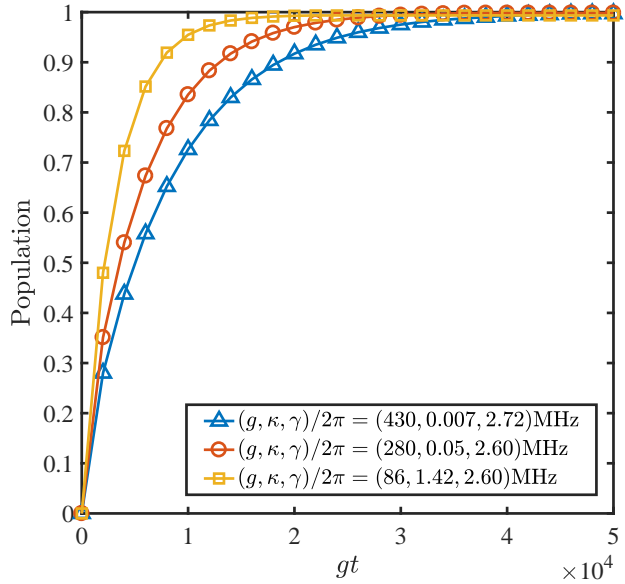}
\caption{\label{ex} The populations of $|S\rangle$ as functions of $gt$ with different experimental parameters. The detuning parameters of microwave field are $\delta=0.01g$. The other relations are $\Omega=0.01g,\Omega_{MW}=0.55\Omega,$ and $J=g$. The initial states are all $|00\rangle|0\rangle_f$. }
\end{figure}

{Finally, we discuss the experimental feasibility of our scheme, and consider the decay of cavities and fiber equal}. In toroidal microresonators with coupling coefficient $g/2\pi=430$ MHz, the cavity field and atomic decay rate are chosen as $(\kappa,\gamma)/2\pi=(0.007,2.72)$ MHz \cite{pra013817}. Based on the parameters, the population of $|S\rangle$ can arrive at $99.99\%$. And then, performing the parameters of silica microspheres $(g,\kappa,\gamma)/2\pi=(280,0.05,2.60)$ MHz \cite{pra013817}, the population of $|S\rangle$ is acquired as $99.98\%$. In \cite{pra013817}, another cavity quantum electrodynamics parameters of $(g,\kappa,\gamma)/2\pi=(86, 1.42, 2.60)$ MHz are calculated. Using the parameters, the populations of $|S\rangle$ can reach $99.48\%$. In Fig.~\ref{ex}, we plot the time evolution of populations for $|S\rangle$ with above parameters, where the representations demonstrate the practicability of our scheme to realize a high-fidelity distant entanglement.

\subsection{For the bipartite KLM state}
\begin{figure*}
\begin{minipage}[t]{0.32\linewidth}
\centering
\includegraphics[scale=0.32]{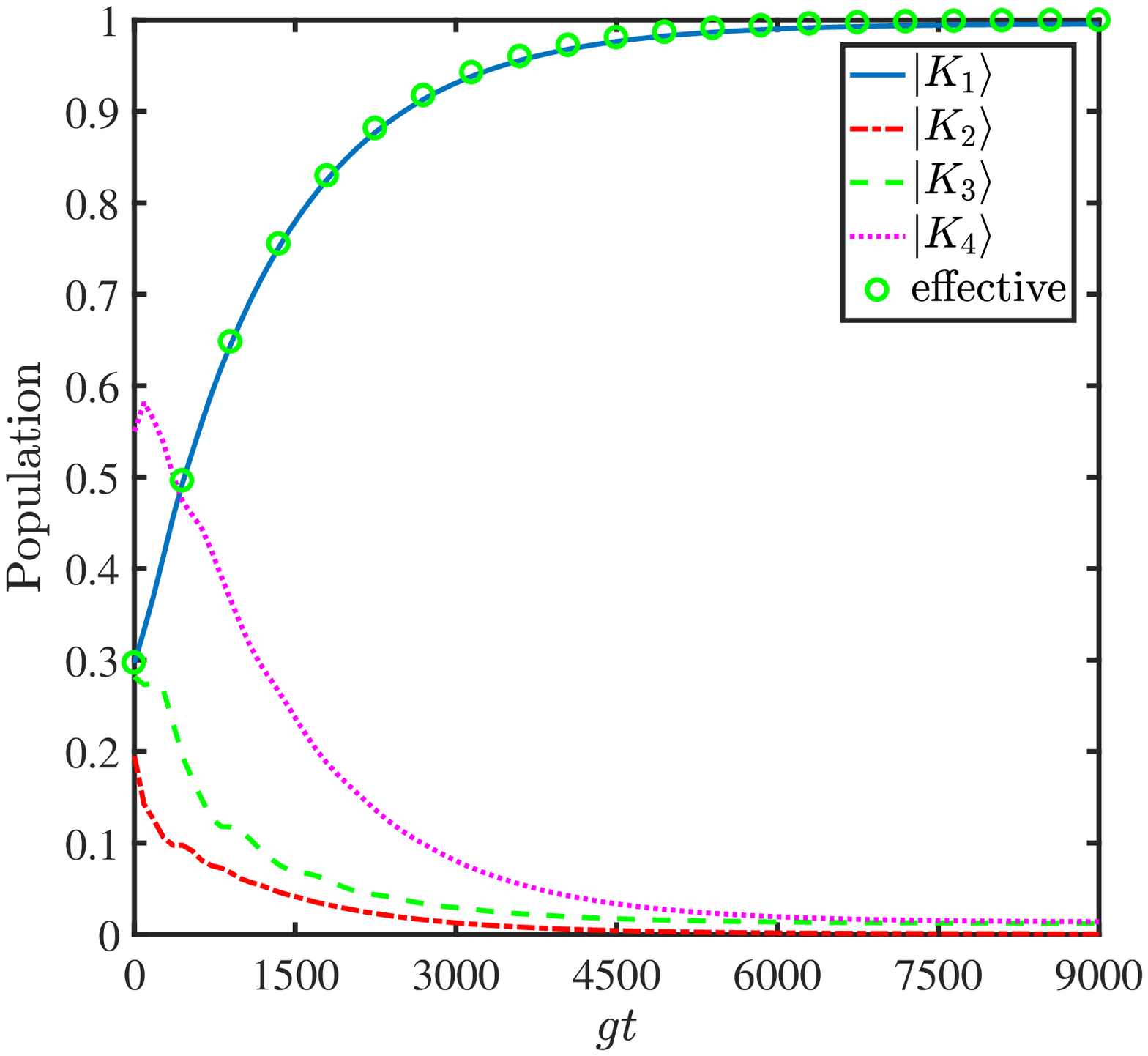}
 \centerline{(a)}
\end{minipage}
\begin{minipage}[t]{0.32\linewidth}
\centering
\includegraphics[scale=0.32]{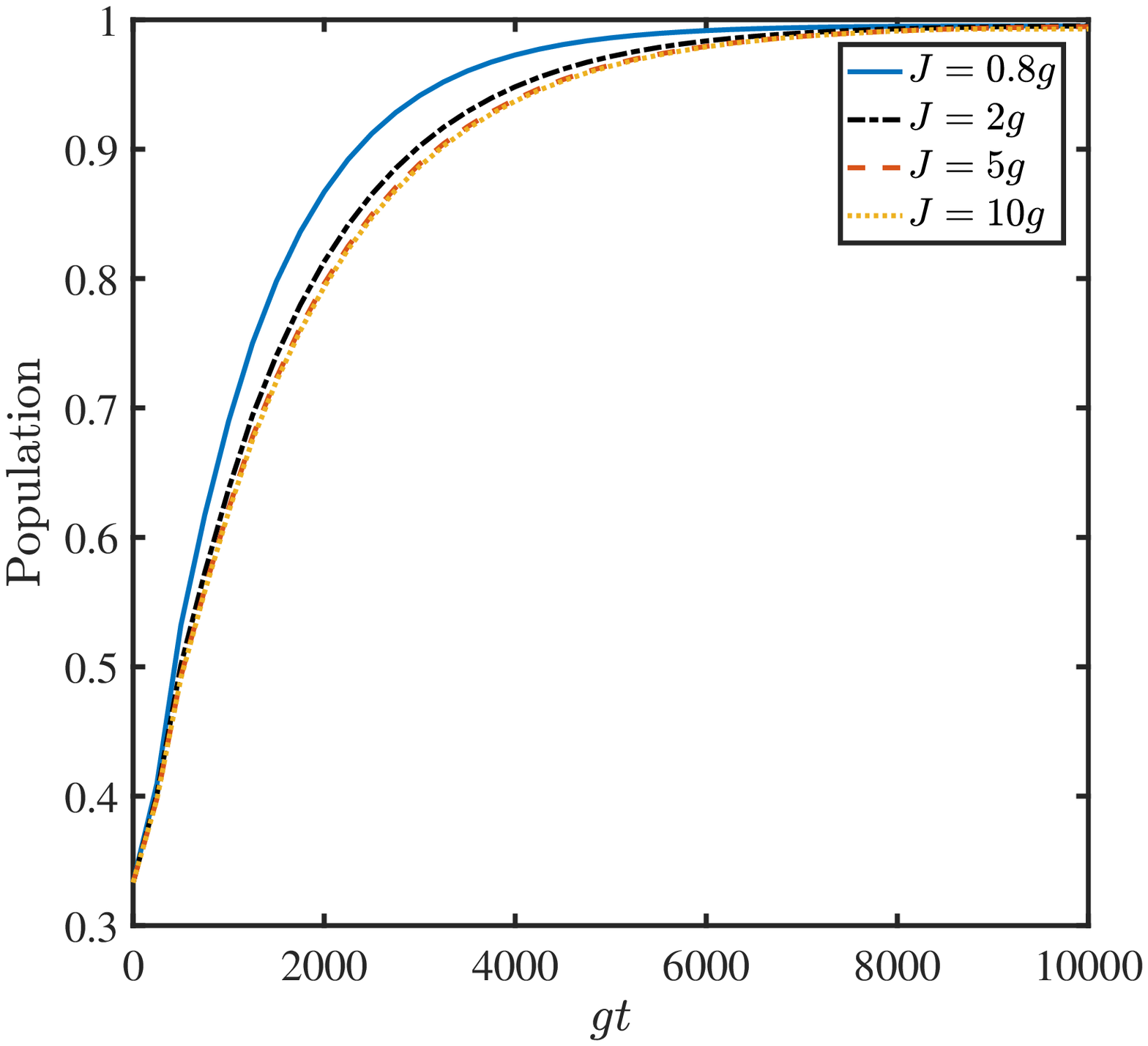}
 \centerline{(b)}
\end{minipage}
\begin{minipage}[t]{0.32\linewidth}
\centering
\includegraphics[scale=0.32]{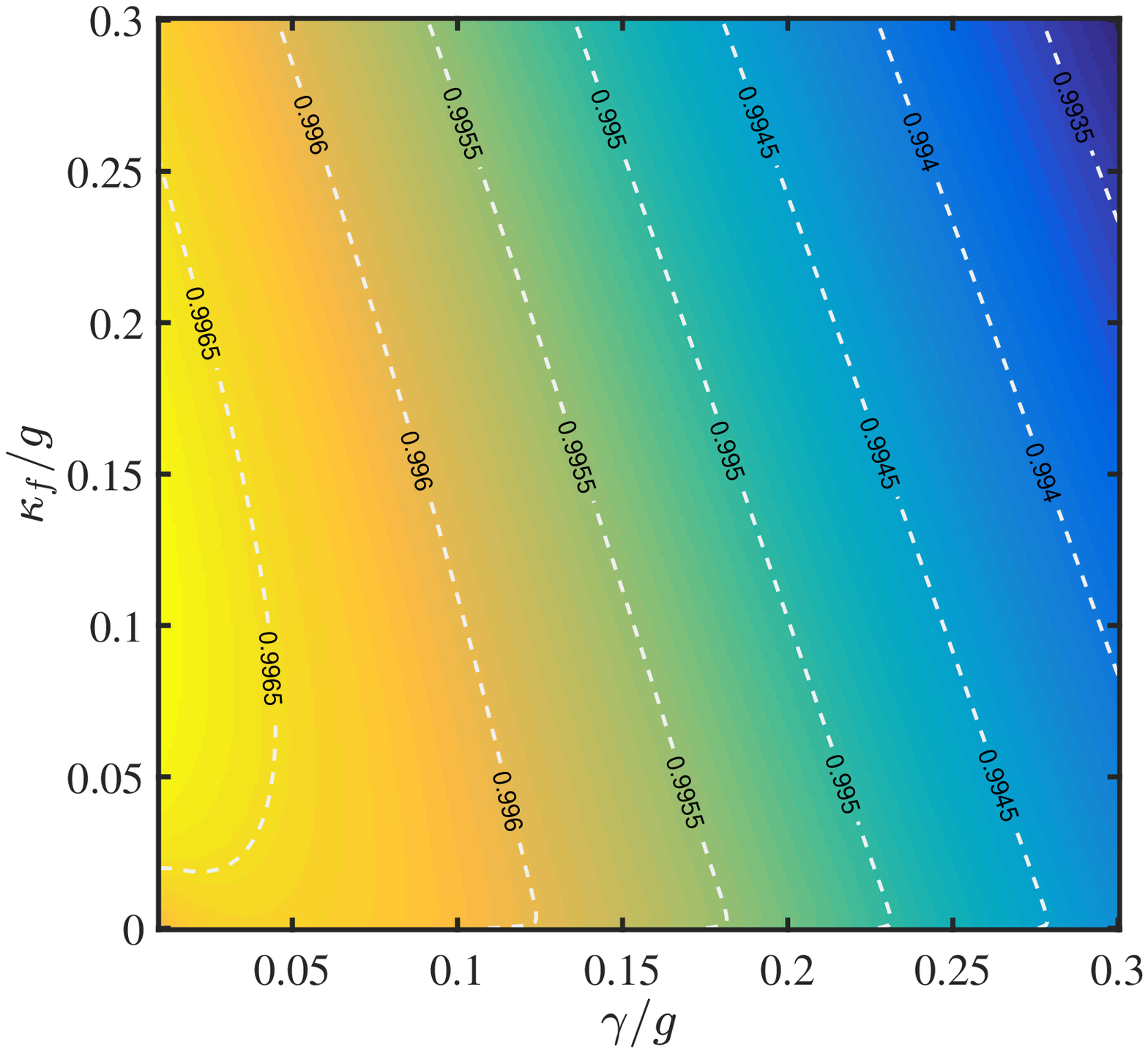}
 \centerline{(c)}
\end{minipage}
\caption{\label{Kfulleffect} (a) The populations as functions of $gt$ governed by full and effective master equations. The initial states are $\rho_0=(a|00\rangle\langle00|+b|11\rangle\langle11|+c|10\rangle\langle10|+d|01\rangle\langle01|)\otimes|0\rangle_f\langle0|$, where $a=0.12$, $b=0.42$, $c=0.35$, and $d=0.11$. We set $\gamma=\kappa_f=0.1g$ and $J=g$. (b) The populations of state $|K_1\rangle$ with different coupling strengths between the fiber and
the cavity $J$. The initial states are chosen as $|00\rangle|0\rangle_f$ and set $\gamma=\kappa_f=0.1g$. (c) Contour plot (dashed lines) of the populations of $|K_1\rangle$ in the steady states with perfect cavities and $J=g$. The other parameter are all set as $\Omega=0.05g,\Omega_{MW}=0.1\Omega$ and $\kappa_1=\kappa_2=0$.}
\end{figure*}
{ In Fig.~\ref{Kfulleffect}(a), we show the populations of $|K_1\rangle$ (solid line), $|K_2\rangle$ (dotted-dashed line), $|K_3\rangle$ (dashed line) and $|K_4\rangle$ (dotted line) governed by the full master equation, where
\begin{eqnarray}
|K_2\rangle&=&\frac{1}{\sqrt{15}}(|00\rangle-3|01\rangle-2|10\rangle+|11\rangle)|0\rangle_f,\\
|K_3\rangle&=&\frac{1}{\sqrt{5}}(\theta_+|00\rangle+|01\rangle-|10\rangle-\theta_-|11\rangle)|0\rangle_f,\\
|K_4\rangle&=&\frac{1}{\sqrt{5}}(\theta_-|00\rangle-|01\rangle+|10\rangle-\theta_+|11\rangle)|0\rangle_f,
\end{eqnarray}
and $\theta_\pm=(\sqrt{5}\pm1)/\sqrt{2}$. It sufficiently attests the feasibility of our proposal that the population of $K_1$ can reach $99.6\%$ While $gt=9000$. The correctness of the reduced system can be also proved by the good agreement between the solid line and the empty circle, which denotes the population of $|K_1\rangle$ governed by the effective master equation.

In Fig.~\ref{Kfulleffect}(b), we also discuss the effect of coupling strength $J$ and the quality of populations is unaffected by coupling strength.  The result reflects the feature, that the scheme need not precisely tailored Rabi frequencies or coupling strength between cavity and fiber, again. The coupling strength only impacts on the convergence time. The stronger $J$ is given, the longer convergence time, which is opposite to it in Fig.~(\ref{fulleffect})(b). The appearance can be also explained by the effective transitions of Eq.~(\ref{klm3CHeff1}) and the Eq.~(\ref{klm3CLind}). When the coupling strength increases, the transition from $|D\rangle$ to $|11\rangle$ will be remarkably inhibited by the decreasing of $\gamma_1$, although the $\gamma_{2,3}$ and $\Omega'/\sqrt{2}$ will be enhanced slightly.

In order to investigate the robustness of the scheme against the decay of fibers and atoms, we plot the populations of $|K_1\rangle$ in the steady states with perfect cavities as functions of $\kappa_f/g$ and $\gamma/g$ in Fig.~\ref{Kfulleffect}(c). The dashed lines are the  contours of the populations and we can find that when $\gamma=0.3g$ and $\kappa_f=0.3g$, the population is still above $99.3\%$. But there is a little decreasing when the decay of fiber and atoms tend to zero. The reason can be obtained by the derivation of effective Lindblad operators in Appendix \ref{B}. The transition from $|D\rangle$ to $|11\rangle$, which is an important resource to realize the bipartite KLM state, consists of the decay of fiber and atoms. So while the two processes are weak, the quality of the target state will decrease.

\begin{figure*}
\begin{minipage}[t]{0.49\linewidth}
\centering
\includegraphics[scale=0.43]{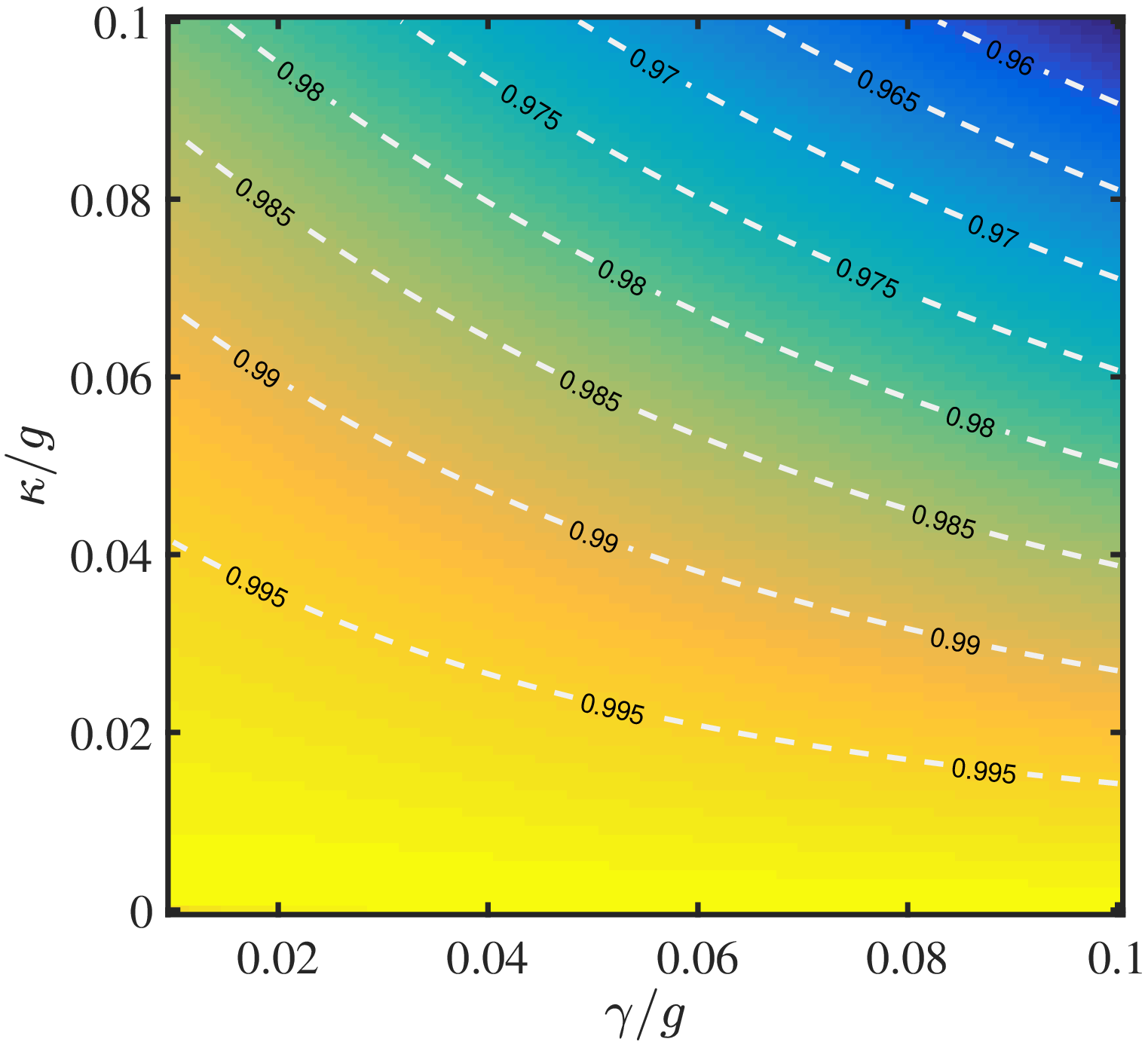}
 \centerline{(a)}
\end{minipage}
\begin{minipage}[t]{0.49\linewidth}
\centering
\includegraphics[scale=0.43]{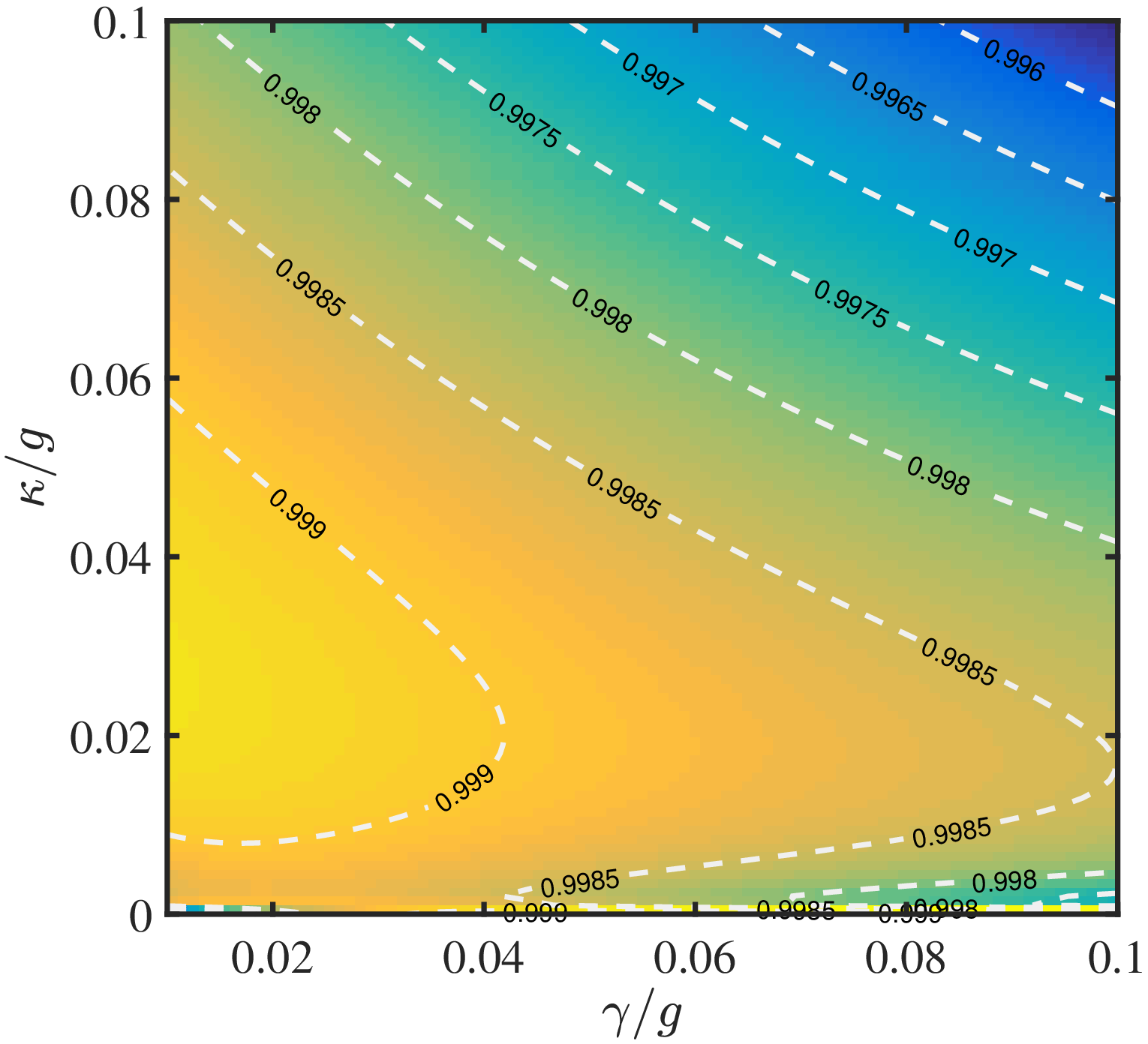}
 \centerline{(b)}
 \end{minipage}
\caption{\label{kfeedback} (a) Contour plot (dashed lines) of the populations of $|K_1\rangle$ in the steady states without the feedback control. (b) Contour plot (dashed lines) of the populations of $|K_1\rangle$ in the steady states based on the feedback master equation with the detection of the two cavities. The feedback parameter is $\eta=0.5\pi$.   The other parameters are all chosen as $\Omega = 0.01g,\Omega_{MW} = 0.3\Omega,J = g,$ and $\kappa_1=\kappa_2=\kappa_f=\kappa$. }
\end{figure*}

\begin{figure}
\centering
\includegraphics[scale=0.51]{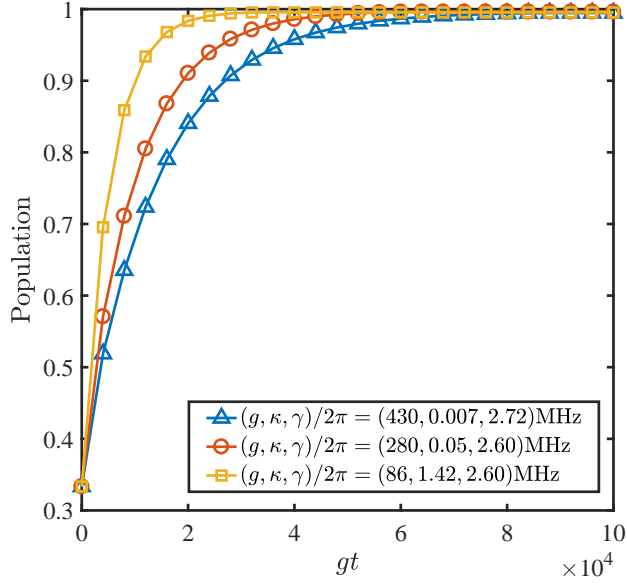}
\caption{\label{kex} The populations of $|K_1\rangle$ as functions of $gt$ with different experimental parameters. The parameters are $\Omega=0.01g,\Omega_{MW}=0.55\Omega,$ and $J=g$. The initial states are all $|00\rangle|0\rangle_f$.}
\end{figure}

In Figs.~\ref{kfeedback}(a) and \ref{kfeedback}(b), we consider the atoms in two leaky cavities without and with the feedback control and plot the populations of $|K_1\rangle$ in the steady states, respectively. In the absence of the quantum-jump-based feedback control, the populations can still arrive at $95.5\%$ even if the decay of cavities, fiber and atoms is $0.1g$. In Fig.~\ref{kfeedback}(b), we introduce the feedback operator $U_{\rm fb}^k=\exp{[ -i\eta(|1\rangle_1\langle0|-|0\rangle_1\langle1|)\otimes {\rm I}_2 ]}$ and consider the feedback control act on the
first atom immediately after detecting the decay of the two cavities. The feedback master equation can be obtained as
\begin{eqnarray}
\dot\rho=-i[H_I,\rho]+(\mathcal{L}_d+\kappa\sum_{i=1}^2\mathcal{D}[U_{\rm fb}^ka_i])\rho.
\end{eqnarray}
After utilizing the quantum-jump-based feedback control, when the atomic spontaneous emission, the decay of cavities and fiber all arrive at $0.1g$, the population of the bipartite KLM state is yet above $99.5\%$, which demonstrate the experimental operability of our scheme. To further reflect the experimental feasibility, we plot Fig.~\ref{kex} with the experimental parameters the same as those in Fig.~\ref{ex} and consider the decay of cavities and fiber equal. The populations of the three groups, $(g,\kappa,\gamma)/2\pi=(430,0.007,2.72)$ MHz, $(g,\kappa,\gamma)/2\pi=(280,0.05,2.60)$ MHz and $(g,\kappa,\gamma)/2\pi=(86,1.42,2.60)$ MHz, can reach $99.58\%$, $99.77\%$ and $99.67\%$, respectively, and all show a high-fidelity
distant entanglement can be achieved by the scheme.}

\section{Generalization: distant entanglement in multi-cavity system}\label{IV}
\subsection{Multi-cavity system of the Bell state}\label{IV1}

Based on the atomic levels in Fig.~\ref{model}(a), we consider there are $n$ cavities connected by $(n-1)$ fibers, $n=2,3,4\ldots$. For each fiber, only one mode is resonant with the cavity mode. The two atoms are trapped into the first and $n$th cavities, respectively. The setup is shown in Fig.~\ref{nmodel}. In the interaction picture, the total Hamiltonian reads
\begin{eqnarray}\label{mutiH}
H_I^{n}&=&H_C^I+H_Q^I,\\
H^I_C&=&\sum_{i=1,n}\Omega_{MW}|1\rangle_i\langle0|+{\rm H.c.}+\delta|1\rangle_i\langle1|-\sum_{j=1}^{n-1}\delta b_j^\dag b_j-\sum_{k=1}^n\delta a_k^\dag a_k,\nonumber\\
H_Q^I&=&\sum_{j=1}^{n-1}Jb_{j}^\dag(a_j+a_{j+1})+\sum_{i=1,n}a_i|2\rangle_i\langle1|g+\Omega_i|2\rangle_i\langle0|+{\rm H.c.}\nonumber
\end{eqnarray}
where $\Omega_1=(-1)^{n}\Omega_{n}$, $a_j(a_j^\dag)$ denotes the annihilation (creation) operator of the $j$th quantized cavity. The Lindblad operators are
\begin{eqnarray}\label{multLind}
L_\gamma^{1(2)}&=&\sqrt{\frac{\gamma}{2}}|0(1)\rangle_1\langle2|,\  \  L_\gamma^{3(4)}=\sqrt{\frac{\gamma}{2}}|0(1)\rangle_{n}\langle2|,\\
L_\kappa^{k}&=&\sqrt{\kappa_k}a_k,\  \  L_{\kappa_f}^j=\sqrt{\kappa_f^j}b_j.
\end{eqnarray}

\begin{figure*}
\centering
\includegraphics[scale=0.21]{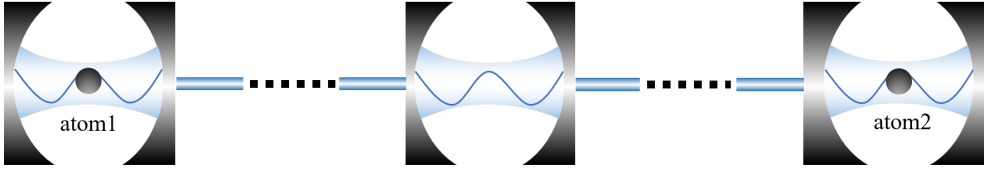}
\caption{\label{nmodel} The model of $n$ cavities connected by $(n-1)$ fibers, $n=2,3,4\ldots$. The two atoms are trapped into the first and $n$th cavities, respectively, and atomic levels are same as the Fig.~\ref{model}(a).}
\end{figure*}

For simplify, we only take into account that $\gamma=\kappa_f^j$. $\kappa_f^j,\kappa_k$ are the decay rates of the $j$th fiber and $k$th cavity, respectively. In the limiting condition $g\gg\Omega$, the effective Hamiltonian for the model can be written as
\begin{eqnarray}\label{mutiHeff}
H_{\rm{eff}}^{n}&=&\Omega'|D\rangle\langle T|+\sqrt{2}\Omega_{MW}|T\rangle(\langle00|+\langle11|){}_f\langle{\bf0}|+{\rm H.c.}\nonumber\\
&&+\delta(2|11\rangle\langle11|\otimes|{\bf0}\rangle_f\langle{\bf0}|+|T\rangle\langle T|+|S\rangle\langle S|+|D\rangle\langle D|),
\end{eqnarray}
where $\Omega'=\sqrt{2}\Omega/\sqrt{G_n+2}$, $G_n=(n-1)(g/J)^2,$ and
\begin{eqnarray}\label{multstate}
|D\rangle&=&\frac{1}{\sqrt{G_n+2}}\Big(|21\rangle|{\bf0}\rangle_f+(-1)^{n}|12\rangle|{\bf0}\rangle_f+\frac{g}{J}\sum_{j=1}^{n-1}(-1)^{j}|11\rangle|{\bf1_j}\rangle_f\Big),\\
|T\rangle&=&\frac{1}{\sqrt{2}}(|01\rangle+|10\rangle)|{\bf0}\rangle_f,\\
|S\rangle&=&\frac{1}{\sqrt{2}}(|01\rangle-|10\rangle)|{\bf0}\rangle_f,
\end{eqnarray}
where {$|{\bf0}\rangle_f=|0\dots0_j\dots0\rangle_f$ represents that all of fibers are in vacuum states and $|{\bf1_j}\rangle_f=|0\ldots01_j0\ldots0\rangle_f$} means that the $j$th fiber stays in the single photon state while other fibers are in vacuum state.

The effective Lindblad operators are denoted as
\begin{eqnarray}\label{multLind}
\ L_{\rm{eff}}^{1}=\gamma_1|11\rangle|{\bf0}\rangle_f\langle D|,\  \  L_{\rm{eff}}^{2(3)}=\gamma_{2(3)}|T(S)\rangle\langle D|,
\end{eqnarray}
where $\gamma_1=\sqrt{\gamma(1+G_n)/(G_n+2)}$ and $\gamma_{2(3)}=\sqrt{\gamma/(2G_n+4)}$.

\begin{figure*}
\begin{minipage}[t]{0.49\linewidth}
\centering
\includegraphics[scale=0.43]{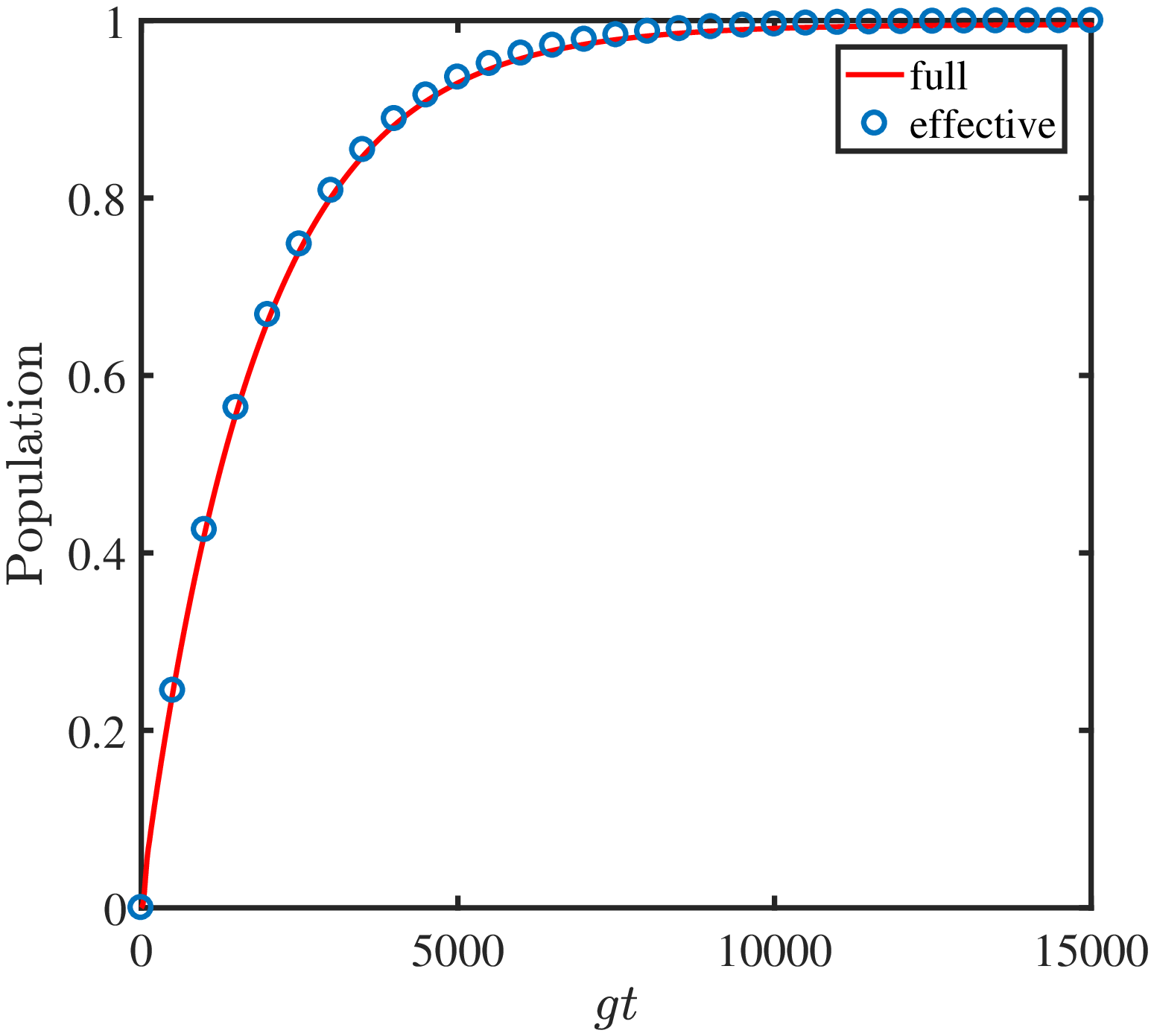}
 \centerline{(a)}
\end{minipage}
\begin{minipage}[t]{0.49\linewidth}
\centering
\includegraphics[scale=0.43]{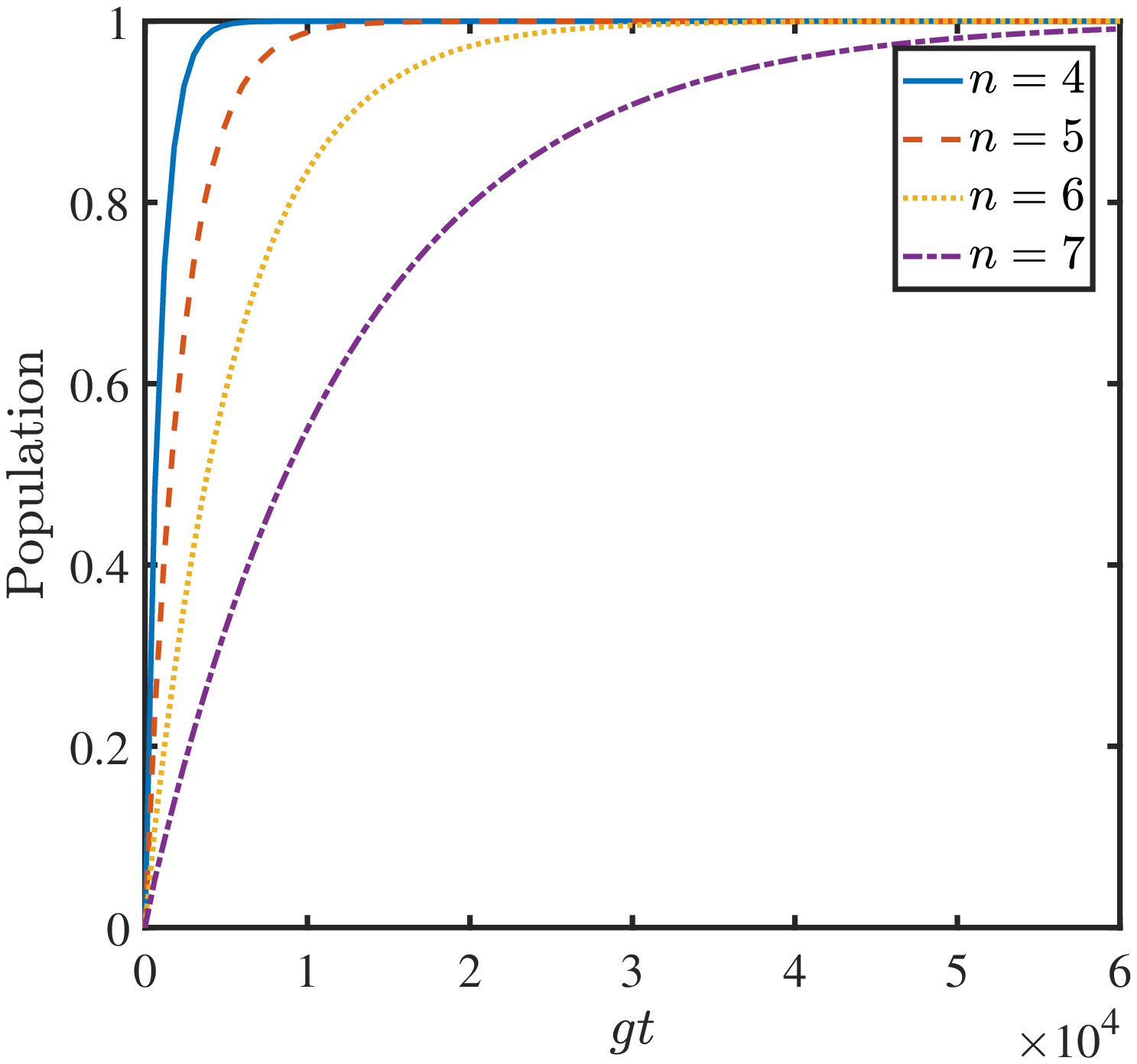}
 \centerline{(b)}
\end{minipage}
\caption{\label{multi} (a) The populations of $|S\rangle$ governed by full (dashed lines) and effective (empty circles) master equations as functions of $gt$. We consider $n=3,\Omega=0.05g,\kappa_f^1=\kappa_f^2=\gamma=0.1g,$ and $\kappa_1=\kappa_2=\kappa_3=0$. The detuning parameters and Rabi frequencies of microwave fields are set as $\Omega_{MW}=0.3\Omega$ and $\delta=0.05g$. (b) The populations of $|S\rangle$ governed by  effective master equations as functions of $gt$ with different $n$. We make use of $\Omega=g/5\cos{[{(n-1)\pi}/({2n})]},J=g,\kappa_f^j=\gamma=0.1g,j=1,2,\cdots,n-1,\Omega_{MW}=0.3\Omega,$ and $\delta=0.05g$.}
\end{figure*}
We compare the time evolution of populations of state $|S\rangle$ governed by the full and the effective master equations with $n=3$ in Fig.~\ref{multi}(a). The initial state is state $|00\rangle|{\bf0}\rangle_f$. The population can be also at $99.52\%$. Due to the achievement of limiting condition of the quantum Zeno dynamics, it has a good approximation for the behaviors of populations for target states calculated by the full and the effective master equations, which reflect the validity of the reduced system again. Then we plot the time evolution of populations of state $|S\rangle$ governed by effective master equation with different $n$ in Fig.~\ref{multi}(b).  Combining the Appendix \ref{A}, we have
\begin{eqnarray}
|\eta_n|_{\rm min}=2\cos{[{(n-1)\pi}/({2n})]},
\end{eqnarray}
which is the minimal absolute value of non-zero eigenvalues of $H=\sum_{j=1}^{n-1}Jb_{j}^\dag(a_j+a_{j+1})+\sum_{i=1,n}a_i|2\rangle_i\langle1|g+{\rm H.c.}$  in the subspace of single excitation. With the change of $n$, we must numerically guarantee the limit condition of quantum Zeno dynamics at all time. Therefore, we assume $\Omega=g|\eta_n|_{\rm min}/10$.  From the Fig.~\ref{multi}(b), the increasing of the number of cavity prolongs the convergence time of generating target states, which can be interpreted by the $\Omega',\gamma_1,\gamma_2$ and $\gamma_3$ in the Eq.~(\ref{mutiHeff}) and the  Eq.~(\ref{multLind}). The enhancement of $n$ will hinder $|D\rangle$ translating to $|T\rangle$ and $|S\rangle$, but promote the decay from $|D\rangle$ to the undesired state $|11\rangle$. However, the final result of state $|S\rangle$ is not upset and the population of state $|S\rangle$ is still high enough.

\subsection{Multi-cavity system of the bipartite KLM state}
\begin{figure*}
\begin{minipage}[t]{0.49\linewidth}
\centering
\includegraphics[scale=0.43]{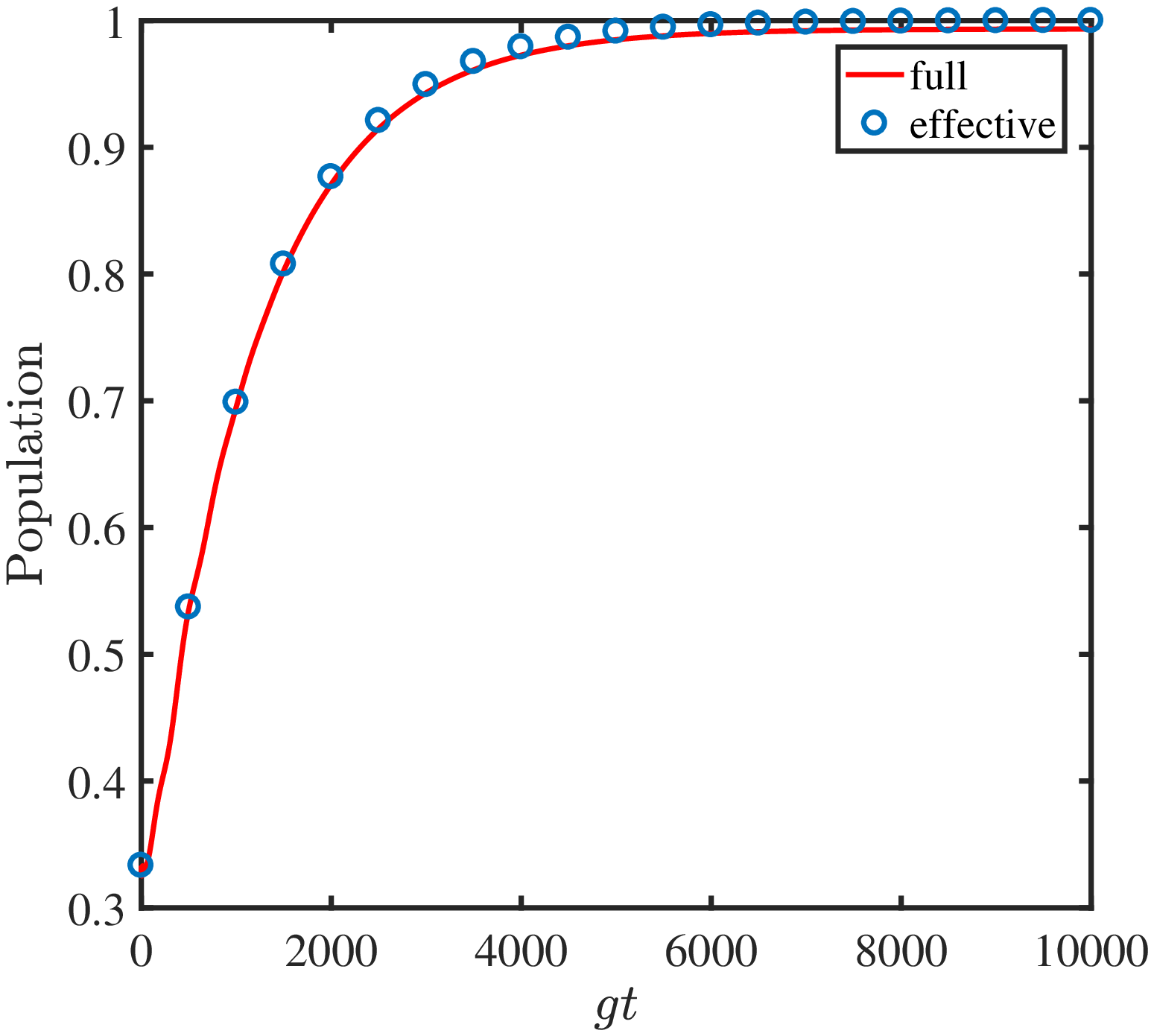}
 \centerline{(a)}
\end{minipage}
\begin{minipage}[t]{0.49\linewidth}
\centering
\includegraphics[scale=0.43]{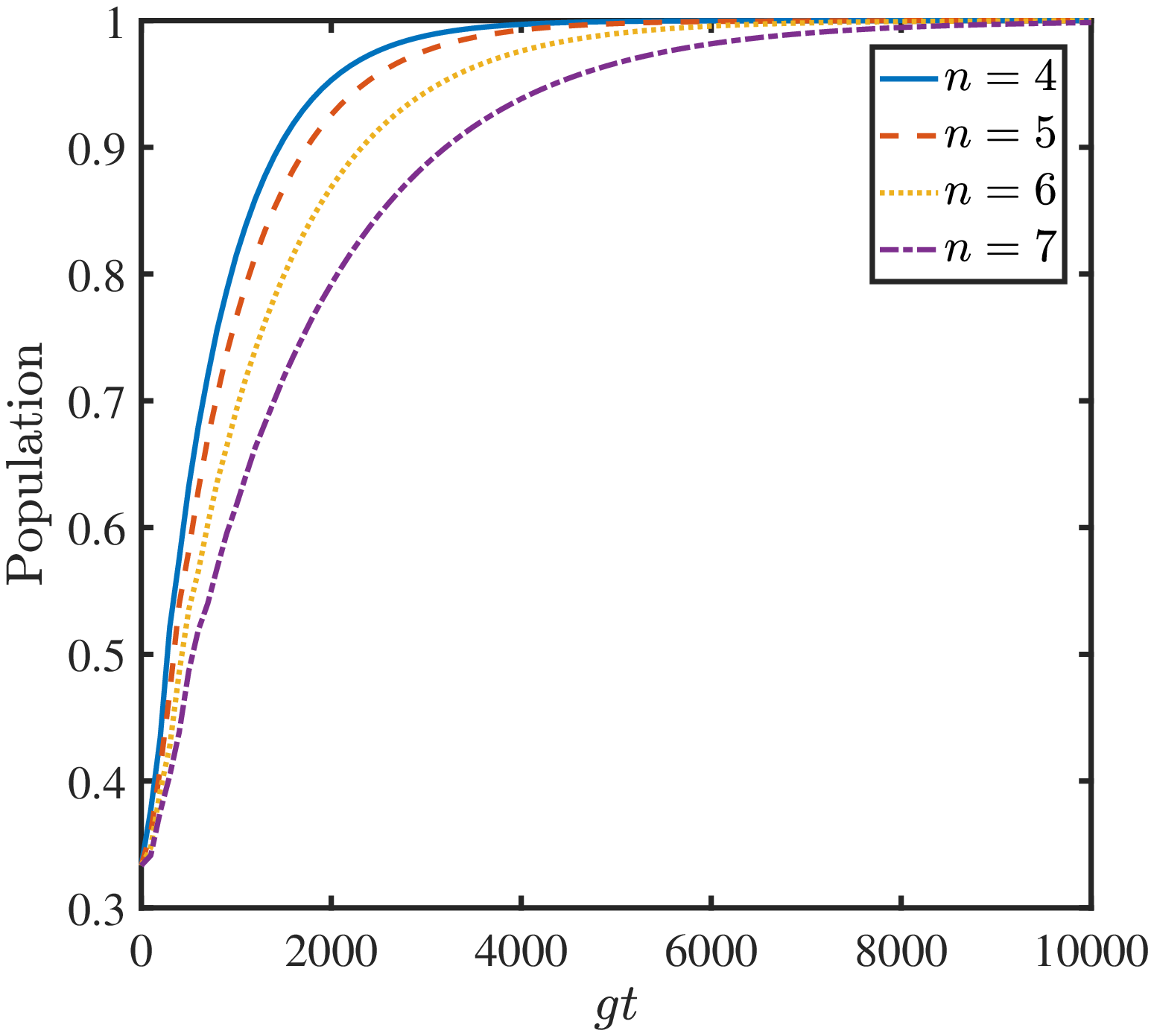}
 \centerline{(b)}
\end{minipage}
\caption{\label{kmulti} (a) The populations of $|K_1\rangle$ governed by full (dashed lines) and effective (empty circles) master equations as functions of $gt$. We consider $n=3,\Omega=0.05g,\kappa_f^1=\kappa_f^2=\gamma=0.1g,\kappa_1=\kappa_2=\kappa_3=0$ and $\Omega_{MW}=0.1\Omega$. (b) The populations of $|K_1\rangle$ governed by effective master equations as functions of $gt$ with different $n$. We make use of $\Omega=g/5\cos{[{(n-1)\pi}/({2n})]},J=g,\kappa_f^j=\gamma=0.1g,j=1,2,\cdots,n-1,$ and $\Omega_{MW}=0.1\Omega$. The initial states are all $|00\rangle|{\bf 0}\rangle_f$.}
\end{figure*}
{ Introducing the atomic levels of Fig.~\ref{klmmodel}(a) into Fig.~\ref{nmodel} and only considering one mode of each fiber
resonant with the cavity mode, we can obtain the Hamiltonian of multi-cavity system to prepare the bipartite KLM state, $|K_1\rangle=(|00\rangle+|10\rangle+|11\rangle)|{\bf 0}\rangle_f/\sqrt{3}$.
\begin{eqnarray}\label{klmmutiH}
H_I^{n}&=&H_C^I+H_Q^I,\\
H^I_C&=&\Omega_{MW}(|1\rangle_1\langle0|-|1\rangle_{n}\langle0|)+{\rm H.c.}+\sum_{i=1,n}\delta|1\rangle_i\langle1|-\sum_{\alpha=1}^{n-1}\delta b_\alpha^\dag b_\alpha-\delta\sum_{\beta}^na_\beta^\dag a_\beta,\nonumber\\
H_Q^I&=&\sum_{\alpha=1}^{n-1}Jb_{\alpha}^\dag (a_\alpha+a_{\alpha+1})+\sum_{i=1,n}a_i|2\rangle_i\langle1|g+\Omega_1|2\rangle_1\langle0|+{\rm H.c.}.\nonumber
\end{eqnarray}
At the same approximations as those in Sec.~\ref{IV1}, the effective Hamiltonian reads
\begin{eqnarray}\label{klmmutiHeff}
H_{\rm{eff}}^{n}&=&\frac{\Omega'}{\sqrt{2}}|01\rangle|{\bf 0}\rangle_f\langle D|+\Omega_{MW}(|00\rangle-|11\rangle)(\langle10|-\langle01|)\otimes|{\bf 0}\rangle_f\langle{\bf 0}|+{\rm H.c.}\nonumber\\
&&+\delta(2|11\rangle\langle11|+|01\rangle\langle 01|+|10\rangle\langle 10|)\otimes|{\bf 0}\rangle_f\langle{\bf 0}|+\delta|D\rangle\langle D|,
\end{eqnarray}
where $\Omega'=\sqrt{2}\Omega/\sqrt{G_n+2}$, $G_n=(n-1)(g/J)^2,$ and the effective Lindblad operators are
\begin{eqnarray}\label{klmmultLind}
\ L_{\rm{eff}}^{1}=\gamma_1|11\rangle|{\bf0}\rangle_f\langle D|,L_{\rm{eff}}^{2(3)}=\gamma_{2(3)}|01(10)\rangle|{\bf0}\rangle_f\langle D|,
\end{eqnarray}
where $\gamma_1=\sqrt{\gamma(1+G_n)/(G_n+2)}$ and $\gamma_{2(3)}=\sqrt{\gamma/(2G_n+4)}$.

We plot the time evolution of populations of state $|K_1\rangle$ governed by the full and the effective master equations in the situation $n=3$ in Fig.~\ref{kmulti}(a). The solid line is in good agreement with empty circles and the population can reach $99.33\%$, which certify the validity of the reduced multi-cavity system and the feasibility of the full multi-cavity system, respectively. In Fig.~\ref{kmulti}(b), we also plot the populations of state $|K_1\rangle$ governed by effective master equation with different $n$ and illustrate the bipartite KLM state of high quality can be generated in a multi-cavity system.}

\section{Summary}\label{V}
In summary, we have systematically discussed the feasibility for generating distant entangled Bell state and KLM state of two atoms with two separated optical cavities. In our analysis, the dispersive microwave field accurately chooses the desired states and the correctness of reduced system are precisely confirmed. The robustness against the decoherence of system are twofold: (i) the atomic spontaneous emission and fiber decay are propitious to generating the target states. (ii) the cavity decay are combated efficaciously by the quantum Zeno dynamics and quantum-jump-based feedback control technology. Ultimately, we extend the two cavities into $n$ cavities connected by $(n-1)$ fibers. The high-fidelity realizations of distant entanglement by the generalized schemes are also investigated. We believe that the schemes supply new prospects to prepare distant entanglement.

\appendix
\section{Derivation of effective Hamiltonian}\label{A}
In this appendix, we show the detailed derivation of effective Hamiltonian. In the situation $J\geq g$, the explicit expression of Eq.~(\ref{limit}) is
\begin{eqnarray}\label{HQ}
H_Q^I&=&\Omega (H_c+KH),\\
H_c&=&\sum_{i=1}^2|2\rangle_i\langle0|+{\rm H.c.},\nonumber\\
H&=&\sum_{i=1}^2a_i|2\rangle_i\langle1|+\frac{J}{g}a_i^\dag b+{\rm H.c.},\nonumber
\end{eqnarray}
where $K=g/\Omega$. Then extending the Eq.~(\ref{HQ}) by the eigenprojections of $H$, we have
\begin{eqnarray}\label{HQ2}
H_Q^I=\Omega \left(\sum_{n}K\eta_nP_n+\sum_{m}P_mH_c\sum_{n}P_n\right),
\end{eqnarray}
where $P_n$ is the eigenprojection of $H$ corresponding to eigenvalue $\eta_n$.
While the limiting condition satisfies $\min{\{K(\eta_m-\eta_n)\}}\gg1$, \textit{i.e.} $\min{\{g(\eta_m-\eta_n)\}}\gg\Omega$, the Eq.~(\ref{HQ2}) can be simplified as
\begin{eqnarray}\label{ApHQ}
H_Q^I=\sum_{n}\Omega K\eta_nP_n+\Omega P_nH_cP_n.
\end{eqnarray}
Because the initial states are chosen in the subspace of $\eta_0=0$ ({the cavities are in the vacuum state}), the Eq.~(\ref{ApHQ}) develops into
\begin{eqnarray}\label{ApHQfinal}
H_Q^I=\Omega P_0H_cP_0=\frac{\sqrt{2}\Omega}{\sqrt{G_2+2}}|D\rangle\langle T|+{\rm H.c.},
\end{eqnarray}
where the subspace of {$P_0$ consists of $\{|D\rangle$, $|10\rangle|0\rangle_f$, $|01\rangle|0\rangle_f$, $|11\rangle|0\rangle_f$, $|00\rangle|0\rangle_f\}\otimes|0\rangle_c$} corresponding $\eta_0=0$. The result is the same as Eq.~(\ref{HQIDT}).

{On the other hand, while $J<g$, the $H$ and $K$ of Eq.~(\ref{HQ}) will be transformed as
\begin{eqnarray}\label{KH}
H=\sum_{i=1}^2\frac{g}{J}a_i|2\rangle_i\langle1|+a_i^\dag b+{\rm H.c.},\  \
K=\frac{J}{\Omega},
\end{eqnarray}
and the limiting condition turns out to be $\min{\{J(\eta_m-\eta_n)\}}\gg\Omega$. The other processes are similar to the situation of $J\geq g$ and the ultimate effective form of $H_Q^I$ is equal to Eq.~(\ref{ApHQfinal}). }

Finally, expanding $H_C^I$ in Eq.~(\ref{B3CH}) by $|D\rangle|0\rangle_c,|T\rangle|0\rangle_c,|S\rangle|0\rangle_c,|11\rangle|0\rangle_f|0\rangle_c,|00\rangle|0\rangle_f|0\rangle_c$ and combining Eq.~(\ref{ApHQfinal}), the effective Hamiltonian of Eq.~(\ref{B3CHeff}) can be derived ({since the cavities are decoupled to our interested system, we may neglect the symbol of the vacuum state $|0\rangle_c$ for simplicity}).

\section{Derivation of effective Lindblad operators}\label{B}

{ The Derivations of effective Lindblad operators of Bell state and KLM state are similar. So we only show the detail of deriving the effective Lindblad operators of Bell state.}

The Lindblad operators of the full master equation are
\begin{eqnarray}\label{App3CLind}
L_\gamma^{1(2)}&=&\sqrt{\frac{\gamma}{2}}|0(1)\rangle_1\langle2|,\  \  L_\gamma^{3(4)}=\sqrt{\frac{\gamma}{2}}|0(1)\rangle_2\langle2|,\  \
L_{\kappa_f}=\sqrt{\kappa_f}b,
\end{eqnarray}
where we have omitted the decay of cavities because the reduced system has been decoupled to cavity field.

Then we project the Lindblad operators into the Hilbert space constructed by $\{ |D\rangle,|T\rangle,|S\rangle,|11\rangle|0\rangle_f,|00\rangle|0\rangle_f  \}$ and consider $\gamma=\kappa_f$,
\begin{eqnarray}
&&L=\sum_{\alpha,\beta}|\alpha\rangle\langle \alpha|L|\beta\rangle\langle \beta|,\alpha,\beta\in\{ |D\rangle,|T\rangle,|S\rangle,|11\rangle|0\rangle_f,|00\rangle|0\rangle_f   \}.
\end{eqnarray}
Then we have
\begin{eqnarray}\label{ApLrho}
L^{1}_\gamma&=&\sqrt{\frac{\gamma}{4G_2+8}}(|T\rangle+|S\rangle)\langle D|,\  \
L^{2}_\gamma=\sqrt{\frac{\gamma}{2G_2+4}}|11\rangle|0\rangle_f\langle D|,\\
L^{3}_\gamma&=&\sqrt{\frac{\gamma}{4G_2+8}}(|T\rangle-|S\rangle)\langle D|,\  \
L^{4}_\gamma=\sqrt{\frac{\gamma}{2G_2+4}}|11\rangle|0\rangle_f\langle D|,\\
L_{\kappa_f}&=&\frac{g}{J}\sqrt{\frac{\gamma}{G_2+2}}|11\rangle|0\rangle_f\langle D|.
\end{eqnarray}

Hence the Lindblad term of full master equation can be rewritten as
\begin{eqnarray}
\mathcal{L}\rho=\mathcal{L}^1\rho+\mathcal{L}^2\rho+\mathcal{L}^3\rho+\mathcal{L}^4\rho+\mathcal{L}^5\rho,
\end{eqnarray}
where
\begin{eqnarray}
\mathcal{L}^1\rho&=&\frac{\gamma}{4G_2+8}(|T\rangle+|S\rangle)\langle D|\rho|D\rangle(\langle T|+\langle S|)-\frac{\gamma}{4G_2+8}(|D\rangle\langle D|\rho+\rho|D\rangle\langle D|),\\
\mathcal{L}^3\rho&=&\frac{\gamma}{4G_2+8}(|T\rangle-|S\rangle)\langle D|\rho|D\rangle(\langle T|-\langle S|)-\frac{\gamma}{4G_2+8}(|D\rangle\langle D|\rho+\rho|D\rangle\langle D|),\\
\mathcal{L}^{2(4)}\rho&=&\frac{\gamma}{2G_2+4}|11\rangle|0\rangle_f\langle D|\rho|D\rangle\langle11|{}_f\langle0|\-\frac{\gamma}{4G_2+8}(|D\rangle\langle D|\rho+\rho|D\rangle\langle D|),\\
\mathcal{L}^5\rho&=&\frac{\gamma G_2}{G_2+2}|11\rangle|0\rangle_f\langle D|\rho|D\rangle\langle11|{}_f\langle0|-\frac{\gamma G_2}{2G_2+4}(|D\rangle\langle D|\rho+\rho|D\rangle\langle D|).
\end{eqnarray}
Furthermore,
\begin{eqnarray}
\mathcal{L}\rho=\mathcal{L}^1_{\rm eff}\rho+\mathcal{L}^2_{\rm eff}\rho+\mathcal{L}^3_{\rm eff}\rho,
\end{eqnarray}
where
\begin{eqnarray}
\mathcal{L}^1_{\rm eff}\rho&=&\frac{\gamma(G_2+1)}{G_2+2}|11\rangle|0\rangle_f\langle D|\rho|D\rangle\langle11|{}_f\langle0|-\frac{\gamma(G_2+1)}{2G_2+4}(|D\rangle\langle D|\rho+\rho|D\rangle\langle D|),\\
\mathcal{L}^2_{\rm eff}\rho&=&\frac{\gamma}{2G_2+4}|S\rangle\langle D|\rho|D\rangle\langle S|-\frac{\gamma}{4G_2+8}(|D\rangle\langle D|\rho+\rho|D\rangle\langle D|),\\
\mathcal{L}^3_{\rm eff}\rho&=&\frac{\gamma}{2G_2+4}|T\rangle\langle D|\rho|D\rangle\langle T|-\frac{\gamma}{4G_2+8}(|D\rangle\langle D|\rho+\rho|D\rangle\langle D|).
\end{eqnarray}
The effective Lindblad operators can be obtained as
\begin{equation}
L_{\rm{eff}}^{1}=\sqrt{\frac{\gamma(G_2+1)}{(G_2+2)}}|11\rangle|0\rangle_f\langle D|,\  \  \
L_{\rm{eff}}^{2(3)}=\sqrt{\frac{\gamma}{2G_2+4}}|S(T)\rangle\langle D|,
\end{equation}
which correspond to Eq.~(\ref{B3CLind}).

\section*{Funding}
National Natural Science Foundation of China (NSFC) (11534002, 61475033, 11774047); Fundamental Research Funds for the Central Universities (2412016KJ004).

\bibliographystyle{apsrev4-1}
\bibliography{MuCavnew}

\end{document}